\documentclass[journal]{IEEEtran}
\usepackage{cite}
\usepackage{amsmath,amssymb,amsfonts}
\usepackage[inline]{enumitem}
\usepackage{color}
\usepackage{subcaption}
\usepackage{graphicx}
\usepackage{xspace}
\usepackage{booktabs}
\usepackage{bm}
\usepackage{multirow}
\usepackage{footmisc}
\usepackage{soul,comment}

\usepackage{cleveref}
\usepackage{textcomp}
\usepackage{dblfloatfix}
\usepackage{tabularx}
\newcommand{\sysname}{AICP\xspace}
\usepackage[hyphens]{url}
\usepackage[linesnumbered,ruled,vlined]{algorithm2e}
\newcommand{\fakeparagraph}[1]{\vspace{.5em}\noindent\textbf{#1}.\hspace{0.25em}}

\newcommand{\vb}[1]{\textcolor{black}{#1}}
\let\oldnl\nl
\newcommand{\nonl}{\renewcommand{\nl}{\let\nl\oldnl}}

\newcommand{\pengyuan}[1]{\textcolor{black}{#1}}

\newcommand{\benjamin}[1]{\textcolor{black}{#1}}

\usepackage{array}
\newcolumntype{L}[1]{>{\raggedright\let\newline\\\arraybackslash\hspace{0pt}}m{#1}}
\newcolumntype{C}[1]{>{\centering\let\newline\\\arraybackslash\hspace{0pt}}m{#1}}
\newcolumntype{R}[1]{>{\raggedleft\let\newline\\\arraybackslash\hspace{0pt}}m{#1}}

\begin{document}
	\title{AICP: Augmented Informative Cooperative Perception}
	
	\author{
		\IEEEauthorblockN{Pengyuan Zhou\IEEEauthorrefmark{1}, Pranvera Kortoçi\IEEEauthorrefmark{2}, Yui-Pan Yau\IEEEauthorrefmark{3},  Benjamin Finley\IEEEauthorrefmark{2}, Xiujun Wang\IEEEauthorrefmark{4}, Tristan Braud\IEEEauthorrefmark{3}, \\Lik-Hang Lee\IEEEauthorrefmark{5}, Sasu Tarkoma\IEEEauthorrefmark{2}, Jussi Kangasharju\IEEEauthorrefmark{2}, Pan Hui\IEEEauthorrefmark{2,3}}
		\thanks{Pengyuan Zhou is with the Research Center for Data to Cyberspace, University of Science and Technology of China. (email: pyzhou@ustc.edu.cn)}
		\thanks{Pranvera Kortoçi, Benjamin Finley, Sasu Tarkoma and Jussi Kangasharju are with the Department of Computer Science, University of Helsinki, Finland. (email: firstname.lastname@helsinki.fi)}
		\thanks{Yui-Pan Yau, Tristan Braud and Pan Hui are with Hong Kong University of Science and Technology, Hong Kong. Pan Hui is also with University of Helsinki, Finland. (email: arthur.yau@connect.ust.hk, braudt@ust.hk, panhui@ust.hk)}
		\thanks{Xiujun Wang is the with School of Computer Science and Technology, Anhui University of Technology, Anhui, China. (email: wxj@mail.ustc.edu.cn)}
		\thanks{Lik-Hang Lee is with KAIST, S.Korea and University of Oulu, Finland. (email: likhang.lee@kaist.ac.kr)}
		\thanks{Corresponding author: Pengyuan Zhou}
	}

	\markboth{IEEE Transactions on Intelligent Transportation Systems}
	{Shell \MakeLowercase{\textit{Zhou et al.}}: IEEE Transactions on Intelligent Transportation Systems}
	\maketitle
	
	\begin{abstract}
		Connected vehicles, whether equipped with advanced driver-assistance systems or fully autonomous, require human driver supervision and are currently constrained to visual information in their line-of-sight. A cooperative perception system among vehicles increases their situational awareness by extending their perception range. Existing solutions focus on improving perspective transformation and fast information collection. However, such solutions fail to filter out large amounts of less relevant data and thus impose significant network and computation load. \benjamin {Moreover, presenting all this less relevant data can overwhelm the driver and thus actually hinder them.} To address such issues, we present Augmented Informative Cooperative Perception (\sysname), the first fast-filtering system which optimizes the informativeness of shared data at vehicles to improve the fused presentation.
		
		To this end, an informativeness maximization problem is presented for vehicles to select a subset of data to display to their drivers. Specifically, we propose
		\begin{enumerate*}[label=(\roman*)]
			\item a dedicated system design with custom data structure and lightweight routing protocol for convenient data encapsulation, fast interpretation and transmission, and
			\item a comprehensive problem formulation and efficient fitness-based sorting algorithm to select the most valuable data to display at the application layer.
		\end{enumerate*}
		
		We implement a proof-of-concept prototype of \sysname with a bandwidth-hungry, latency-constrained real-life augmented reality application. \pengyuan{The prototype adds only 12.6 milliseconds of latency to a current informativeness-unaware system.} Next, we test the networking performance of~\sysname at scale and show that \sysname effectively filters out less relevant packets and decreases the channel busy time. 
	\end{abstract}
	
	
	\IEEEpeerreviewmaketitle
	
	\section{Introduction}
	\label{sec:introduction}
	Connected and autonomous vehicles are closer than ever to becoming a reality. Specifically, modern communication technologies such as cellular vehicle-to-everything (C-V2X) and dedicated short-range communications (DSRC) facilitate large-scale vehicular communication thanks to significant improvements in bandwidth, latency, and reliability. Additionally, novel regulations provide a beneficial legal context for the operation of autonomous vehicles on public roads\vb{\footnote{\url{https://www.theverge.com/2018/2/26/17054000/self-driving-car-california-dmv-regulations}}}. This paves the way for the deployment of applications that leverage vehicular communication to provide more information to human and AI drivers, thus improving road safety.
	Currently, autonomous vehicles and advanced driver-assistance systems (ADAS) rely heavily on onboard sensors to identify and evaluate potential dangers and take necessary actions. 
	\pengyuan{More specifically, advanced vehicles employ various sensors to observe the environment, a perception module for sensor data fusion~\cite{8959552}, a path planning module for route planning based on previous modules~\cite{7995787}, and a control module for maneuver decision~\cite{8620918}}.
	However, most current solutions are limited to a single-vehicle point of view, sensing only the nearby objects within their line-of-sight. As such, the vehicle's sensing capabilities are regularly obstructed by other vehicles, and thus depriving the driver of potentially useful information.
	Leveraging current and future communication networks, the vehicle can aggregate the perception of multiple nearby vehicles, i.e., cooperative~(collective) perception~\cite{rauch2012car2x,kim2014multivehicle}, and provide a driver (human or AI) with a holistic view of the road situation. This concept has been adopted by the European Telecommunications Standards Institute~(ETSI), which is working on Cooperative/Collective Perception Service standardization~\cite{etsi103324,etsi-tr-103-562}.
	
	\pengyuan{Existing works focus on timely and synchronized information distribution, data fusion, or communication overhead while the \textit{informativeness}\footnote{The timeliness with which we receive given messages as well as the data contained within is strictly related to the capability to identify potential harm-causing objects. We incorporate these notions into the term \emph{informativeness} and use it throughout the rest of the article for the sake of conciseness.} of the shared perception data has been largely overlooked~\cite{rauch2012car2x,kim2014multivehicle,qiu2018avr,garlichs2019generation,zhou2018arve,zhou2019enhanced}}. Cooperative perception, and more generally safety applications that rely on communication among vehicles, require deployment at scale to provide a holistic vision of the road. Such a pervasive deployment leads to a significant strain in terms of network, computation resources, and driver awareness caused by the constant information dissemination across a large number of vehicles. Meanwhile, only part of the disseminated information is of interest to the context of each driver~\cite{a2w}. For example, parking space is more important for drivers in parking lots, while pedestrians require more attention for drivers near intersections.
	
\begin{figure}[!t]
		\centering
		\includegraphics[
		width=.8\linewidth]{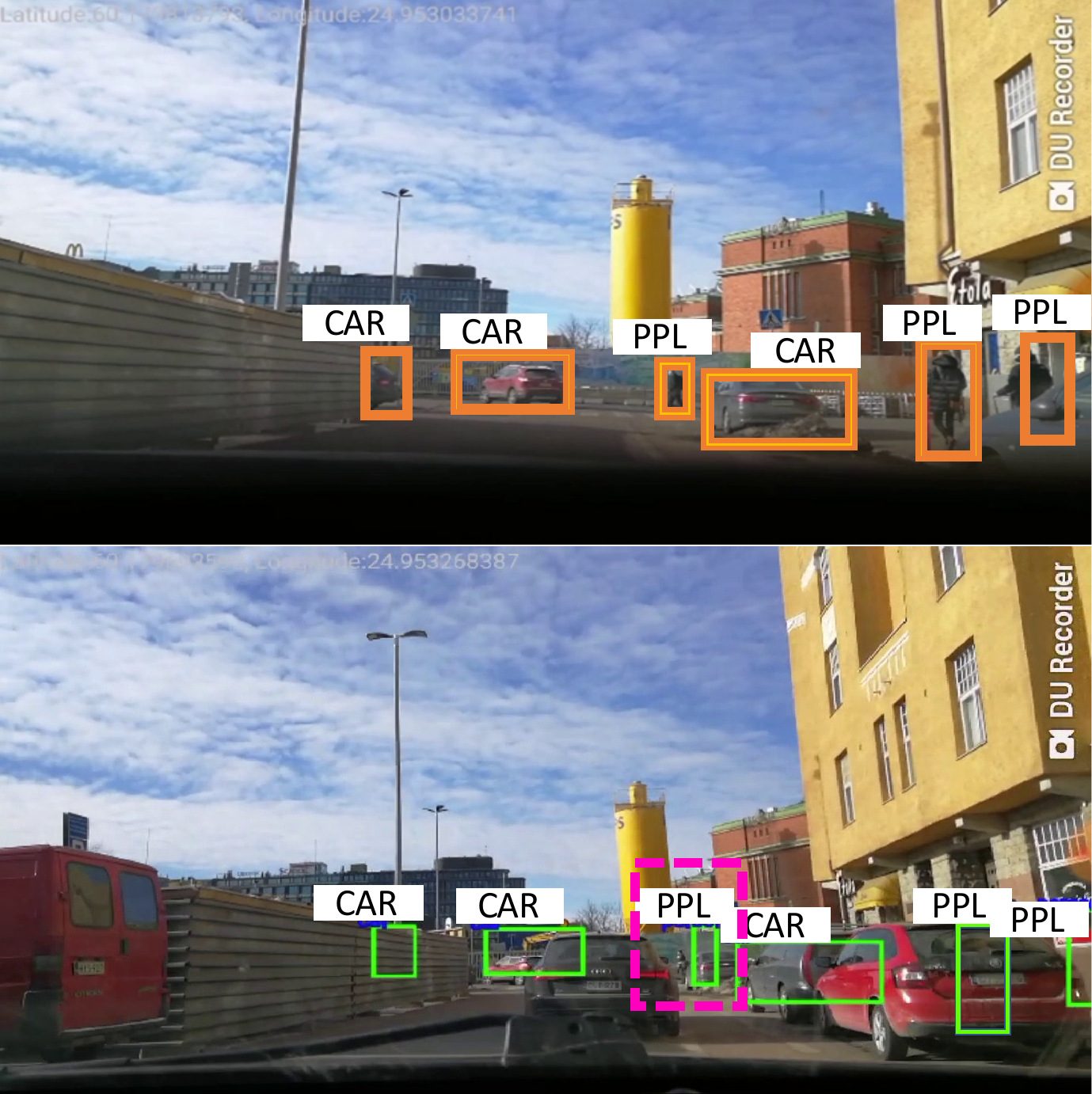}
		\caption{Illustration of a naïve cooperative perception system. A \emph{leading vehicle} detects objects shown in orange bounding boxes (top figure). These objects are displayed to the driver of the \emph{following vehicle} in green bounding boxes (bottom figure). \sysname filters such objects to only show critical ones (e.g., the pedestrians in the pink bounding box) to reduce on-screen object overload.}
		\label{fig:example}
	\end{figure}

	Figure~\ref{fig:example} shows an example of a following vehicle's vision in a leading-following vehicle scenario with naïve cooperative perception with augmented reality (AR). The leading vehicle captures the objects within its line-of-sight and broadcasts information about the detected objects. The following vehicle calculates the position transformations and renders all the objects, shown in green boxes, from the received messages. Extending such a system to \emph{city-block-level} perception with no information filtering leads to a massive number of extraneous objects (given the context) being displayed to the drivers. This overwhelms their vision and thus negatively impacts their driving experience. In fact, a driver's decision time increases logarithmically with the number of stimuli or objects~\cite{liu2020relevant}. Additionally, limiting the number of objects to the human cognition capacity of about 7$\pm$2 items~\cite{Schweickert1986ShorttermMC} is essential. As such, cooperative perception requires an efficient filtering system. For instance, such a system might select the pedestrian shown in the pink box in Figure~\ref{fig:example} as critical and display their information while safely discarding the remaining objects.
	
	The ETSI standards and a few recent works~\cite{garlichs2019generation,thandavarayan2019analysis} have proposed high-level descriptions of potential filtering rules and mechanisms. \vb{However, there are still many topics of dedicated system design that have not been explored. Such is the case of fine-grained protocols that provide efficient data flow with lightweight operations, as well as fast-filtering algorithms that optimize the informativeness of objects in real-time.}
	%
	In this paper, we propose \sysname, the first solution that focuses on optimizing informativeness for pervasive cooperative perception systems with efficient filtering at both the network and application layers. \sysname identifies, transmits, forwards, and filters objects at scale and displays the most informative ones to the drivers through AR. Such an AR-based system will be useful even for vehicles with level four automation (which experts estimate could be at least ten years from wide scale deployment\footnote{\url{https://news.mit.edu/2020/mit-3-questions-john-leonard-future-of-autonomous-vehicles-0804}}), since with this automation level drivers may still need to take vehicle control in some complex situations (as compared to fully autonomous level five).
	Specifically, we make several \textbf{key contributions} as follows.
	
	\vspace*{0.5mm}
	\noindent \textbf{(1)} We propose a system design for~\sysname. The design includes a dedicated data structure, the vehicular data unit~(VDU), designed for informativeness-focused information filtering and transmission. We also describe the full-stack networking protocol, \pengyuan{Contextual Multihop Routing (CMR)}, that the system employs to utilize VDU.
	
	\vspace*{0.5mm}
	\noindent \textbf{(2)}  We formulate the informativeness in cooperative perception systems as a multi-level problem, namely the \textit{object}, \textit{message}, and \textit{vehicle level}, and propose a near real-time sorting algorithm based on Mahalanobis distance for fast yet comprehensive filtering. The algorithm provides filtering at the application level to display only the most important information shared by nearby vehicles and thus preventing drivers from information overload.
	
	\vspace*{0.5mm}
	\noindent \textbf{(3)}  We implement a proof-of-concept (POC) prototype using a cooperative perception application on an Augmented Reality Head-up Display~(ARHUD). We demonstrate the system performance in different contexts via data-driven tests using the data collected with the prototype from road tests. \pengyuan{Next, we evaluate the networking performance of~\sysname at scale. Due to the limit of the testbed, we conducted simulations using a state-of-the-art vehicular network simulator}. 
	
	We note that \sysname is network-agnostic and does not depend on any particular feature of the underlying network. The system can be seamlessly integrated into current and future communication systems such as C-V2X and DSRC.
	
	The rest of the paper is structured as follows. Section~\ref{sec:related_work} discusses related works and states the key motivations behind \sysname. Section~\ref{sec:system} details the system architecture, data structure, and routing protocol. Section~\ref{sec:opt_model} models the system and formulates the problem. Section~\ref{sec:sorting} describes the weighted fitness sorting algorithm to calibrate the assessment of informativeness. Section~\ref{sec:evaluation} shows the POC implementation of~\sysname and its performance in different contexts. Section~\ref{subsec:simulation} presents the simulation setup and results. Finally, Section~\ref{sec:discussion} discusses system limitations and potential solutions, and Section~\ref{sec:conclusion} concludes the work.
	
	
	\section{Related Work}
	\label{sec:related_work}

	This work touches on different research areas including cooperative perception, information filtering, and AR in the context of vehicular networks, and so does the related work.
	
	\vb{Artificial Intelligence (AI) is at the core of many solutions to increase road safety in modern transportation systems including, for example, autonomous vehicles~\cite{ma2020artificial}. Specifically, AI-based perception algorithms help develop an understanding of a vehicle's surrounding environment through architectures such as convolutional and deep neural networks (CNN and DNN) that perform object detection and recognition. In terms of research in this area,~\cite{kebria2019deep} shows that these networks still have significant limitations (that should be considered) and also require optimization of network structure and parameters (e.g., number of layers, filters, and filter sizes) to provide rapid object detection. Similarly,~\cite{cao2020rapid} uses deep learning to rapidly segment blind roads and crosswalks to assist visually-impaired people. Furthermore,~\cite{cao2020rapid} optimizes existing neural network architectures (e.g., by reducing the number of parameters) to speed up the training and inference, while retaining high model accuracy. However, all these solutions suffer from (inherent) high network complexity, and thus require considerable optimization prior to adoption in real-life, latency-sensitive applications.} \benjamin{In comparison, our work does not use a neural network approach but instead relies on basic assumptions about vehicular traffic, and thus does not suffer from this complexity, though as we discuss later neural networks will play a role in our future work.}
	
	In terms of general vehicular cooperative perception, \vb{Yoon et al. presented a decentralized cooperative perception framework~\cite{yoon2021performance}. The work investigates the effects of common inherent limitations of any Vehicle-to-Vehicle (V2V) network, such as the loss of communication and variations in the rate of participating vehicles. The study suggests that the overall perception improves with the vehicle participation rate. However, increasing vehicle participation above a certain \emph{optimal} rate results in no significant perception improvement, though communication and computation costs in the network increase drastically.} Kim et al. proposed a framework addressing several important problems in the field such as map merging, communication uncertainty, and sensor multi-modality~\cite{kim2013cooperative}. Furthermore,~\cite{gunther2016collective} proposed and analyzed different message formats based on ETSI ITS \vb{G5}~\cite{etsi-its-g5-5g} to exchange local sensory data among road participants for collective perception. Thereafter, they proposed a multimodal cooperative perception system with a focus on the engineering feasibility~\cite{kim2014multivehicle}, and generalized the work with a mirror neuron-inspired intention awareness algorithm for cooperative autonomous driving~\cite{kim2016cooperative}. \benjamin{Overall, in contrast, our work focuses on the more pointed problems of information filtering and ranking, which we discuss next, and thus complements these more general works that mainly focus on other problems.}
	
	The critical area of filtering mechanisms in vehicular cooperative perception is still very new and has only a few key works. Garlichs et al.~\cite{garlichs2019generation} in 2019 suggested a set of generation rules to reduce the transmission load while guaranteeing perception capabilities. This proposal was later added to the ETSI standard~\cite{etsi-tr-103-562}. Thandavarayan et al. studied the ETSI standards and conducted an in-depth evaluation of the message generation rules~\cite{thandavarayan2019analysis}. They investigated the trade-off between perception capabilities and communication performance under current standards and concluded that further optimization is needed to reduce information redundancy. To this end, Aoki et al.~\cite{aoki2020} applied a deep reinforcement learning approach to reduce the information sent between vehicles by only forwarding information about objects that are not likely to have been seen directly by surrounding vehicles themselves. \benjamin{Our work differs in that we focus on only forwarding information about objects that are likely to be important (to receiving vehicles for driver awareness) regardless of whether the object is directly seen by surrounding vehicles. Thus the objective is related but broader and in some sense complementary.}
	
	Finally, works in the area of ARHUD, an in-car deployment of AR that visualizes information in the driver's line-of-sight, are also related. For instance,~\cite{qiu2017augmented,qiu2018avr} explore how to share augmented vision between two vehicles. Other studies consider connecting multiple mobile points of view to recompose a scene in 2D or 3D~\cite{chen2019cooper}. \benjamin{However, in comparison with our work, these studies mainly focus on image-stitching, and thus overlook aspects that we focus on such as information importance and filtering.}
	
	To the best of our knowledge, cooperative augmented vehicular vision at scale still requires additional research on efficient data filtering. Specifically, we believe that AR-powered cooperative perception needs a comprehensive solution to maximize the informativeness of the data shared among vehicles to improve the driving experience while increasing road safety. As such, in this work we propose~\sysname, a system that lessens the burden on the network through efficient data filtering. As a result, only the most relevant data is broadcast to vehicles, which in turn sort such data to maximize the informativeness that they yield to the driver.
	
	
	\section{System Overview}
	\label{sec:system}
	\begin{figure}[!t]
		\centering
		\includegraphics[width=.8\columnwidth]{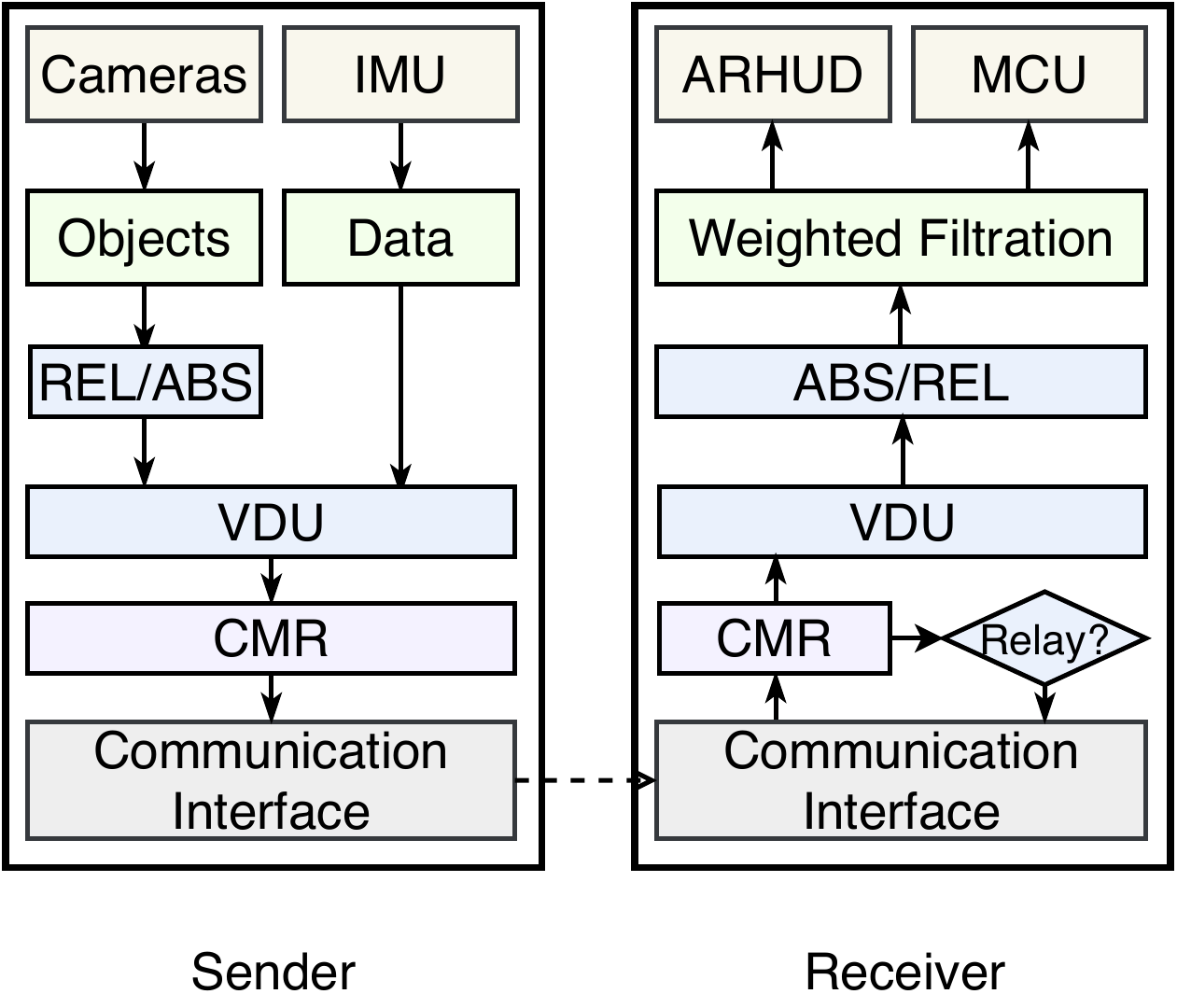}
		\caption{AICP components with information flows. The components include inertial measurement unit (IMU), microprocessor control unit (MCU), relative values (REL), absolute values (ABS), vehicular data unit (VDU), and contextual multihop routing (CMR) protocol.}
		\label{fig:filterflow}
	\end{figure}
	This section describes the proposed system's major components, data structures, and routing protocol.
	
	\subsection{System Architecture}
	\label{subsec:arc}
	We consider a connected system of vehicles equipped with sensors and wireless communication modules that can collect and share sensory data.
	We assume the system has capabilities including accurate positioning and localization~\cite{wolcott2015fast,hata2015feature}, relative velocity estimation, distance and angle estimation~\cite{glenn_jocher_2020_3983579}, and perspective transformation~\cite{qiu2018avr}. In this work, we skip the details of the techniques mentioned above and focus on information filtering.
	Figure~\ref{fig:filterflow} depicts the system architecture including the major data flows between key components. These data flows include:
	\begin{itemize}[wide = 0pt]
		\item IMU sensors in each sender collect sensory data. Each sender detects the objects captured by the onboard cameras and corresponding information such as distance, relative velocity, and moving direction. 
		\item The sender system transforms the object data from relative values to absolute values based on IMU data. For instance, the system transforms the relative velocity of a detected object to absolute velocity by adding its own velocity.
		\item The data are then encapsulated into VDUs~(see Section~\ref{ssec:vdu}).
		\item The sender system encapsulates networking layer information into VDUs according to the CMR~(Section~\ref{ssec:routing}).
		\item The senders and receivers exchange data packets via wireless communication interfaces.
		\item Each receiver decides whether to forward the received packet based on a network layer filter~(see Section~\ref{ssec:routing}).
		\item Each receiver transforms the absolute values of the objects to relative values based on their IMU data.
		\item Each receiver performs object resolution to reconcile data (often from multiple senders) that refers to the same object (seen from different sender perspectives).
		\item Each receiver filters the received VDU based on a filtering algorithm~(see Algorithm~\ref{alg:stack}).
		\item Each receiver ARHUD renders the filtered information through AR to enhance the driver's situational awareness. The MCU performs maneuvers based on the filtered information.
	\end{itemize}
	As such, each vehicle can display the most important situational information in real-time to facilitate safe driving. Next, we describe the VDU and the CMR in detail.
	\begin{figure}[!t]
		\centering
		\includegraphics[width=.7\columnwidth]{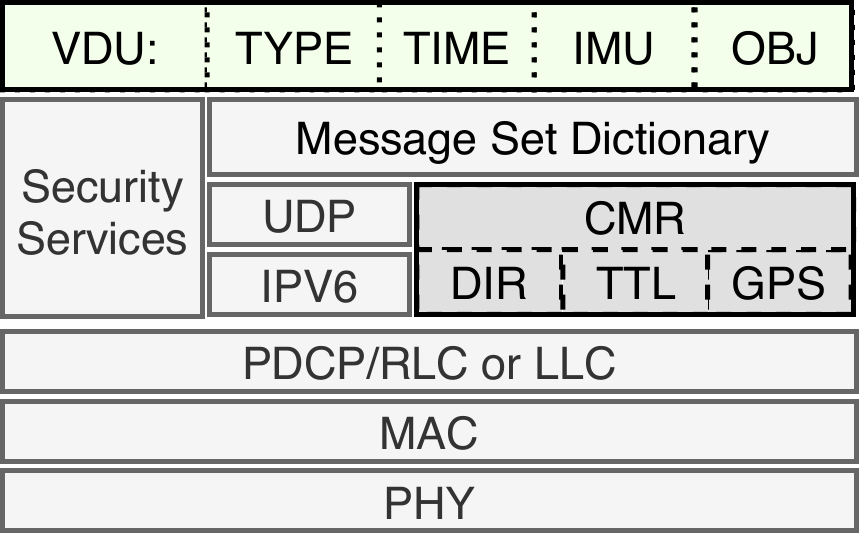}
		\caption{\benjamin{\sysname protocol stack including the contextual mutual routing (CMR) and vehicle data unit (VDU) with their specific fields such as location (GPS) and time (TIME). Other generic network layers, such as the link and MAC layers, are shown for reference.}}
		\label{fig:stack}
	\end{figure}
	
	\subsection{Vehicular Data Unit~(VDU)}
	\label{ssec:vdu}
	Figure~\ref{fig:stack} depicts the overall protocol stack deployed by~\sysname. 
	\sysname is network-agnostic and can be deployed on top of any V2V broadcast-style protocol such as DSRC. CMR is deployed as the routing protocol to provide context-aware routing in a broadcast network environment, in parallel to the traditional UDP/IP stack. Finally, the VDU contains the information required at the application layer.
	To accelerate data processing during filtering, we propose the vehicular data unit~(VDU)
	for vehicular information encapsulation.
	A VDU comprises multiple metadata fields, each of which is a key-value pair or key-value map. The metadata fields include:
	\begin{itemize}[wide = 0pt]
		\item \emph{TYPE} -- whether the message is a safety or non-safety application message (pair).
		\item \emph{TIME} -- the time the information was first created (pair)\footnote{We employ elapsed time since Unix epoch to record timestamps.}.
		\item \emph{IMU} -- the information of IMU sensors (map), i.e., \{\textit{GPS coordinates}, \textit{velocity}, \textit{direction}, \textit{category}\}.
		\item \emph{OBJ} -- the information of detected objects (map).
	\end{itemize}
	Similar to the SAE J2735 standard~\cite{saej2735}, the protocol stack defines a message set dictionary to specify the VDU structure and provides sufficient background information to allow vehicle systems to properly interpret the message. Together with the segmented data blocks, the dictionary extends the system compatibility by allowing different VDU structures. The dictionary and VDU also speed up the look-up process of fields such as \emph{TIME} which can help determine whether the received information is outdated (see Section~\ref{sec:opt_model}).
	
	
	\subsection{Contextual Multihop Routing (CMR)}
	\label{ssec:routing}
	
	Vehicle-to-Vehicle~(V2V) communication suffers from a short communication range due to signal attenuation caused by obstacles like nearby vehicles and buildings~\cite{friedner20165g}. Therefore, packet forwarding (also known as relaying) is crucial to extend the range over which information can be propagated. \benjamin
	{As an example, DSRC has a theoretical range of several hundred meters. However, in common non-line-of-sight situations, like intersections, vehicles potentially cannot communicate over distances as short as 60 meters due to obstacle attenuation over the 5.9 GHz band~\cite{noor2018performance}}.
	
	Additionally, to date, most DSRC and C-V2X standards assume data transmission over a broadcast mechanism. Broadcast transmission allows a vehicle to efficiently forward information to other vehicles in their immediate vicinity. However, broadcast transmission suffers from multiple drawbacks in multihop communication. Unregulated broadcast transmission results in significant data redundancy that affects the system at every level, from increased load and congestion on the transmission medium to large amounts of unnecessary information being forwarded to the drivers. To address these concerns, we introduce CMR as the routing protocol for packet forwarding. 
	
	CMR enables the following features.
	
	\begin{enumerate}[label=(\roman*)]
		\item \textit{Directional routing in a multihop broadcast transmission.} We consider that an object detected by a vehicle is relevant for all immediate neighbor vehicles (accessible within a single hop). \benjamin{In other words, a vehicle that detects an object always forwards the information to neighboring vehicles. However, vehicles that receive such information only process and further forward the information if they are going in the same general direction as the sending vehicle. Specifically, the same general direction is defined as having headings that differ by less than a certain absolute degree threshold (this threshold can vary). This scheme relies on the intuition that vehicles travelling in similar directions will find the same detected objects informative. This represents a trade-off between sending all information and potential congestion (with information loss) or less information (and fewer objects) with less congestion. Though, the scheme can reduce to the former by setting the degree threshold to 180 degrees.}
		
		\benjamin{
			\item \textit{Distance and hop limits for geographic relevance}
			After a certain geographic distance, the information also loses relevance (regardless of vehicle direction). Thus vehicles only forward packets if the geographic distance between the vehicle and the sending vehicle is less than a certain threshold. Additionally, as the number of hops acts as a rough proxy for geographic distance (that does not rely on GPS), a hop limit whereby vehicles only forward packets up to a maximum number of hops also helps enforce this geographic distance. The hop limit also helps remove the impact of potential routing loops.}
		
	\end{enumerate}
	To provide these functions, CMR in this work relies on two different metrics, namely \textit{GPS} and \textit{time-to-live}, respectively referring to the coordinates of the detected object and a counter used to enforce the hop limit (see Figure~\ref{fig:stack}). The GPS coordinates of the detected object are encoded over two fields of $32$ bits to achieve the precision of at least a meter. 
	
	\textbf{Note} that unlike common routing algorithms such as DV-CAST~\cite{tonguz2010dv}, Greedy Perimeter Stateless Routing (GPSR)~\cite{karp2000gpsr} and its variants such as~\cite{lochert2005geographic,granelli2007enhanced}, CMR focuses on filtering low-informativeness packets instead of improving the communication efficiency. Therefore it is a complementary protocol to the efficiency-focused routing protocols instead of a replacement.
	
	For reference, we detail the basic routing algorithm in Algorithm \ref{alg:cmr}. As mentioned, the parameters of the algorithm include the initial $T$ (time-to-live) value, direction threshold, and distance threshold.
	
	\begin{algorithm}[t]
		\SetAlgoLined
		\KwData{Receiver Direction $\textit{DIRR}$, time-to-live $\textit{T}$, Receiver Location $\textit{GPSR}$, Transmitter Direction $\textit{DIRT}$, Transmitter Location $\textit{GPST}$, Direction Threshold $\textit{DT}$, Distance Threshold $\textit{D}$, Packet $\textit{PKT}$}
		
		\textit{T}=\textit{PKT}[\textit{T}]-1\;
		\textit{DIRT}=\textit{PKT}[\textit{DIR}]\;
		\textit{GPST}=\textit{PKT}[\textit{GPS}]\;
		
		\uIf{$\text{T}<0~ \text{or} ~|\text{DIRT}-\text{DIRR}| > DT~ \text{or} ~\text{Distance}(\text{GPST},\text{GPSR}) > D$}{
			drop(PKT)\;
		}
		\Else{
			forward(\textit{PKT})\;
			process(\textit{PKT})\;
		}
		
		\caption{CMR Directional Routing}
		\label{alg:cmr}
	\end{algorithm}

	\section{System Model and Problem Formulation}
	\label{sec:opt_model}
	
	This section details the system model and the problem of maximizing the informativeness of the displayed objects. 
	
	\subsection{System Model}
	\label{ssec:system_model}
	
	The system includes the set $\mathcal{N} = \{ 1, 2, ..., N \}$ of $N$ vehicles driving in the considered area, with velocities  $\mathcal{V}^t = \{ v^t_1, v^t_2, ..., v^t_N \}$ at time $t$, respectively. The mobility of the vehicles is exogenous to the system.
	Each vehicle is equipped with an ADAS or autonomous system consisting of several cameras facing varying directions for comprehensive vision around the vehicle (radar, lidar, and ultrasonic sensors are optional and not a mandatory requirement of \sysname), GNSS/IMU for real-time kinematic and positioning, and wireless interfaces (DSRC or C-V2X) for communications with other devices on the road.
	Messages are sent with a frequency between $1$ and $10$ Hz and the message size is limited to $300$ Bytes, as specified in the C-V2X standard~\cite{gsma-cv2x,etsi-its-g5-5g}.
	The encapsulation and decapsulation of the messages follow the protocol standard defined in Figure~\ref{fig:stack}. We model the data propagation from three parallel levels, namely the \textit{object level}, \textit{message level}, and \textit{vehicle level}. \emph{\sysname} decides whether to display an object based on the \textit{object level} informativeness, to forward a packet based on \textit{message level} informativeness, and targets optimizing performance based on \textit{vehicle level} informativeness.
	
	\begin{table}[t!]
		\renewcommand*{\arraystretch}{1.15}
		\caption{Summary of used notations.}\label{tab:notation}
		\centering
		\begin{tabular}{|c||p{7.05cm}|}
			\specialrule{1.3pt}{1pt}{1pt}
			\textbf{\scalebox{.95}[1.0]{Symbol}} & \textbf{Definition} \\\hline
			$\mathcal{N}$ & Set of all vehicles considered in the system   \\
			$\mathcal{V}^t$ & Set of velocities $\mathcal{V}^t=\{v^t_1, \dots, v^t_N\}$ of $\mathcal{N}$ vehicles \\
			$\textit{T}_{th}$ & Initial T (time-to-live) of detected objects in the system \\  
			$\vartheta^O$ & Informativeness of object $O$  \\
			$t_c$ & Time at which a message is created  \\
			$\mathcal{I}^i(t)$ & Informativeness of message $i$ at time $t$ \\
			$r$ & Decay rate of Informativeness $\mathcal{I}$ over time  \\
			$x^{j,i}(t)$ & Binary variable indicating if message $i$ is received by vehicle $j$ at time $t$  \\
			\specialrule{1.3pt}{1pt}{1pt}
		\end{tabular}
	\end{table}
	
	\fakeparagraph{Object level} 
	The processing result of each image frame contains a list of detected objects, each of which is defined as $O \triangleq \{D, V, R, C\}$, where $D, V, R, C$ denote the \textit{Distance, relative Velocity, diRection and Category} of the object, respectively. The rationale of the choice of these parameters is justified by the fact that an object $O$ has a higher chance of causing an accident if 
	\begin{enumerate*}[label=(\roman*)]
		\item it is close to the vehicle,
		\item is getting closer to the vehicle, e.g., catching up with the vehicle from behind or coming right at the vehicle, and
		\item is on the heading direction of the vehicle. 
	\end{enumerate*}
	Additionally, the rationale for having object categories relies on the fact that certain objects could cause or sustain a greater injury in an accident; e.g., pedestrians and cyclists are more likely to be killed in vehicle accidents compared to vehicle occupants~\cite{peden2004}.
	
	These parameters are intertwined with each other, and such dynamics are crucial to determine a model upon which we define the informativeness of an object. For instance, the mutual time and space relationship are, in fact, at the basis of modern methodologies to evaluate accidents by analyzing the collision area ~\cite{dirnbach2020methodology,higuchi2019value}. An examination of reported accidents involving autonomous vehicles in California showed that most accidents occur at cross sections in suburban roads, with most accidents reporting rear or front damage~\cite{favaro2017examining}. These findings indicate that the direction with which vehicles move greatly affects the probability of an accident, especially if the vehicles are at a short distance from each other. Furthermore, we need to factor in higher informativeness for objects that fall into the \emph{people} category, for instance.
	
	Upon such considerations, we express the informativeness $\mathcal{I}$ of an object $O$ as:
	\begin{equation}\label{eq:object_relevance}
		\pengyuan{\vb{\mathcal{I}}^{O} = f(D, V, R, C)} 
	\end{equation}
	\pengyuan{Since it is not easy to quantify the exact form of function $f$, we instead use a \textit{weighted fitness sorting} algorithm (Section~\ref{ssec:fitness}) to calculate the inter-weight relationships.}
	
	More complicated computations such as machine learning along with accident reports of autonomous vehicles\footnote{\url{https://www.dmv.ca.gov/portal/vehicle-industry-services/autonomous-vehicles/autonomous-vehicle-collision-reports/}}\cite{caesar2020nuscenes} can incorporate the understanding of a chain of scenarios; however, they might suffer from additional costs. We leave the investigation of such an approach as possible future work.
	
	
	\fakeparagraph{Message level} 
	The messages received by each vehicle may arrive at varying times and with varying delays. In fact, due to the highly time-sensitive nature of the warning messages for assisted driving, vehicles need to establish the timeliness of the message~\cite{hannah2020long}. To this end, the system extracts from the VDU the time at which a given message was created (see \textit{TIME} in Figure~\ref{fig:stack}), here denoted as $t_c$, and uses it to evaluate the timeliness. For a message $i$, its informativeness can be calculated as
	\begin{equation}\label{eq:message_timeliness}
		\mathcal{I}^i(t) = 
		\bigg ( \vb{\mathcal{I}}^{i} \Big ( \dfrac{\textit{T}^{i}(t)}{\textit{T}_{th}} \hspace{1mm} (1 - r) \Big ) \bigg )^{(t - t_c^{i})}
	\end{equation}
	where \benjamin{$\vb{\mathcal{I}}^{i}=\max_{o \in O_i} \vb{\mathcal{I}}^{o}$} denotes the informativeness of message $i$ and $O_i$ denotes the detected objects contained in message $i$, $r$ is the rate at which the \textit{informativeness} of the message decays over time, and $t$ denotes the current time. The decay rate $r$ is strictly connected to the time limit within which such messages are considered \emph{up-to-date} and \emph{relevant}. In fact, $r$ is a system parameter that can be tuned for specific conditions. 
	\begin{figure}
		\centering
		\includegraphics[width=.9\linewidth]{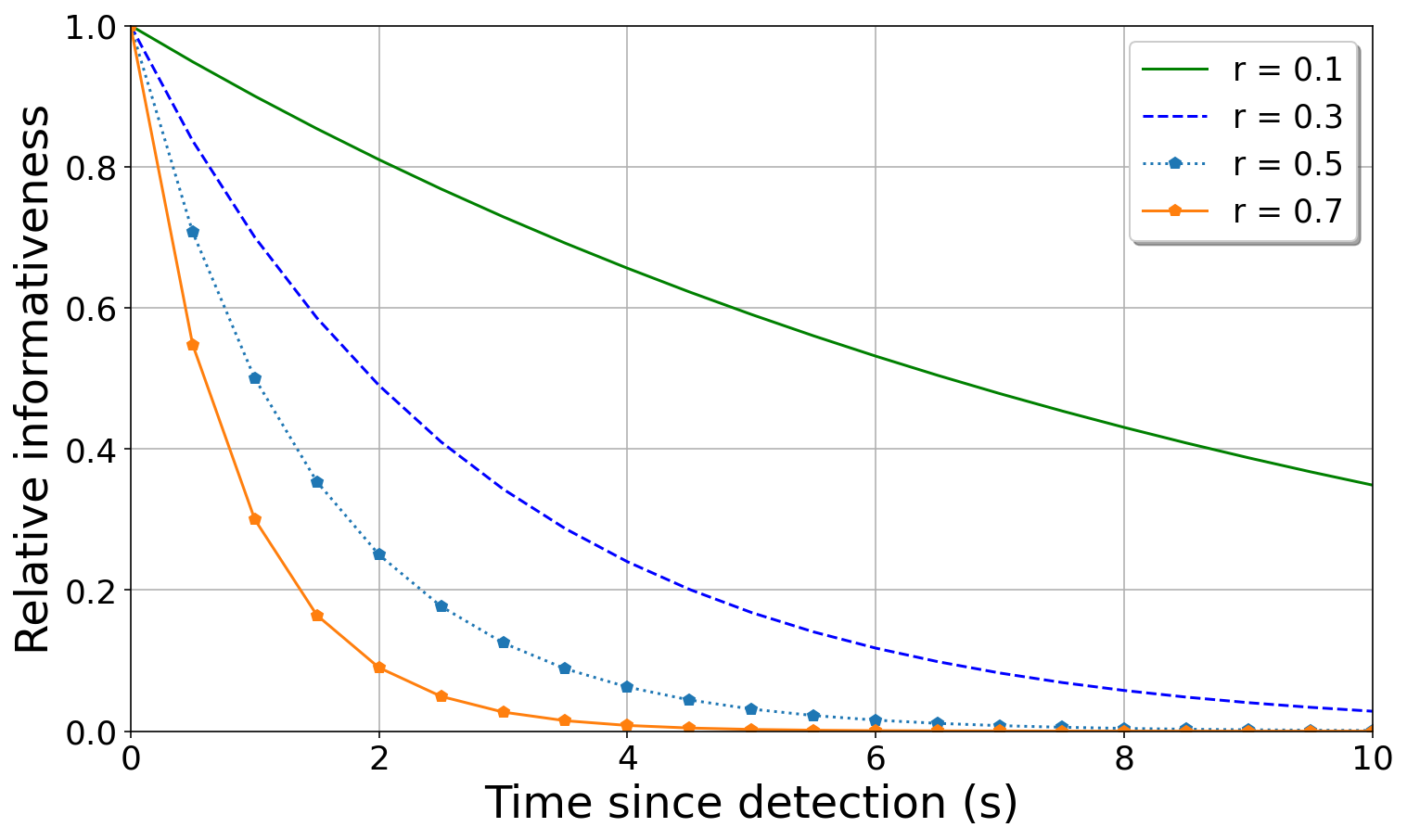}
		\caption{\benjamin{Informativeness of an object relative to its informativeness at detection (t=0) as a function of time elapsed for four different decay rate $r$ parameters.}}
		\label{fig:rel_infor}
	\end{figure}
	\vb{Figure~\ref{fig:rel_infor} shows the informativeness of an object as a function of time elapsed since its time of detection (time = $0$) for different decay rates $r$. We call this parameter \emph{relative informativeness}, and it represents the effect of the decay rate $r$ on the informativeness of an object. This parameter, in fact, dampens an object's informativeness in addition to that already dictated by the \emph{time-to-live} $T$ parameter.}

	\vb{A higher decaying rate is required when road conditions change quickly, such as driving at high speeds on the highway. Similarly, we need to relax (i.e., decrease) the decay rate $r$ for conditions with lower speeds such as driving through a city center. For instance, a decay rate of $r=0.3$ halves the informativeness of a message in about $2$ seconds, whereas a decay rate of $0.1$ requires around $6$ seconds to halve the informativeness of an object. Situations such as that of driving in high traffic require low decay rates; for instance, a car driving as slow as $30$ km/h in an urban area would drive a distance of around $50$ m before the informativeness of its generated messages halves, given a decay rate of $0.1$.} 
	Keep in mind that $r$ can be further tailored should the sampling frequency of the vehicle's sensor(s) vary over time. That is, $r$ increases with the sampling frequency due to the fact that a high sampling frequency could lead to detecting the same object(s) during several consecutive times \pengyuan{and more frequent message exchanges, and thus a vehicle can afford to discard (the same) old messages faster and avoid network congestion}. 
	
	According to the proposed ETSI generation rules of CPMs (collective perception messages)~\cite{etsi-tr-103-562,thandavarayan2020generation}, objects can be generated and broadcast with a frequency between $100$ and $1000$ ms. As such, object(s) can be re-broadcast already several times before their informativeness halves, ensuring a high degree of diffusion for informative objects and thus messages. Moreover, the decay rate $r$ factors in possible redundancy in the exchange of outdated messages (and thus, objects) among vehicles~\cite{thandavarayan2020redundancy}. That is, it serves as a time delimiter for the broadcast of \emph{aged} messages, and thus it ensures that only \emph{fresh} messages are broadcast in the network, similar to~\cite{corneo2019age}. This is especially useful in case there are no or few new detected objects to broadcast, forcing a vehicle to keep broadcasting over time the same outdated messages. As such, the decay rate $r$ serves a dual purpose: first, vehicles transmit and receive only messages (objects) that are fresh and bring a high level of information toward safe driving, and second, it leads to more efficient use of the available channel bandwidth. \benjamin{Relatedly, the use of the max aggregation over object importances when calculating message importance helps ensure that messages with a few important objects but many unimportant objects are not discarded (and thus important objects are not missed).} Finally, ${\textit{T}}_{th}$ characterizes the \textit{hop limit} defined by CMR~(see Section~\ref{ssec:routing}). ${\textit{T}}^{i}(t)$ expresses the remaining time-to-live, i.e., the number of hops a message can still be forwarded, of message $i$ at time~$t$.
	
	\fakeparagraph{Vehicle level} 
	At the vehicle level, a vehicle can
	\begin{enumerate*}[label=(\roman*)]
		\item evaluate a received message as informative and display some objects from that message to the driver and subsequently broadcast the message,
		\item evaluate a received message as irrelevant to themselves but still broadcast the message, or
		\item drop the received message if its information is outdated, and thus not relevant to any vehicle in the network.
	\end{enumerate*}
	
	
	Upon receiving a multitude of messages, a vehicle derives the informativeness of the objects within. We assume a vehicle $n \in \mathcal{N}$ receives $\mathsf{M}$ messages at time $t$ from other vehicles in the network and expresses their informativeness as below. 
	\begin{align}\label{eq:vehicle_message_informativeness}
		\mathcal{I}_{n}(t) = \sum\limits_{\substack{j=1 \\ j \neq n}}^{|\mathcal{N}|} \sum\limits_{\substack{i=1}}^\mathsf{\vb{M_j}} 
		\hspace{-.1cm} \bigg ( \vb{\mathcal{I}}^{j,i}s\Big ( \dfrac{\textit{T}^{i}(t)}{\textit{T}_{th}} \hspace{1mm} (1 - r) \Big ) \bigg )^{(t - t_c^{i,})} x^{j,i}
	\end{align}
	The binary variable $x^{j,i}$ is $1$ if 
	message $i$ is received by vehicle $j$ ($x^{j,i} = 1$), and $0$ otherwise.
	
	\vb{The number of received messages (and objects) can increase drastically when a vehicle is in dense traffic, thus displaying all objects would overwhelm the drivers' field of view and cognitive processing. As such, we next present a mechanism to filter incoming messages by optimizing (i.e., maximizing) their informativeness.}
	
	
	
	\subsection{Problem Formulation}
	\label{ssec:problem}
	
	The following formulates an optimization problem for a vehicle to select objects whose informativeness helps to identify imminent risks and thus increase road safety.
	
	Specifically, the problem is defined as below.
	\begin{equation}\label{eq:opt_prob_obj}
		\max_{j,i} \hspace{.2cm} \mathcal{I}_{n}(t)
	\end{equation}
	Subject to:
	
	\allowdisplaybreaks
	\begin{align}
		\label{utility:constraint:obj_limit}
		& \qquad \sum\limits_{\substack{j=1 \\ j \neq n}}^{|\mathcal{N}|} \sum\limits_{\substack{i=1 }}^\mathsf{\vb{M_j}} 
		x^{j,i} \leq L, \hspace{.2cm} \forall n \in \mathcal{N} \\
		\label{utility:constraint:obj_ttl}
		& \qquad \textit{T}^{i}(t) > 0, \hspace{1.6cm}  \forall i \\
		\label{utility:constraint:integer}
		& \qquad  x^{i,j} \in \{0,1\}, \\ & \forall j \in [1, 2, ..., |\mathcal{N}|], \forall i \in [1, 2, ..., \mathsf{M}]
	\end{align}
	
	The objective function in Eq.~\eqref{eq:opt_prob_obj}, defined as the summation of the informativeness of the objects contained in the messages incoming from other vehicles, expresses the \emph{collective informativeness} that vehicle $n$ conveys at time~$t$. Eq.~\eqref{utility:constraint:obj_limit} limits the number of selected objects used to convey information to the driver. This allows us to display only clear and limited audio and visual content to a driver~\cite{correia2017avui}. Finally, Eq.~\eqref{utility:constraint:obj_ttl} specifies that the time-to-live of an object must be, trivially, larger than $0$. 
	Given a suitable value of $L$, \sysname selects $L$ objects with the largest informativeness 
	to display to the driver. 
	
	We also note that we assume that object resolution (also known as entity resolution) is performed before this informativeness selection step. Object resolution is the process of recognizing that multiple perceived or received objects are actually the same object seen by multiple vehicles at different angles. As this resolution process is a well-known topic in itself, we do not focus further on the process but instead refer to~\cite{papadakis2020blocking}, which summarizes different object resolution methods. These methods typically have super-linear but sub-quadratic time complexity with the number of objects and thus should not significantly impact the system performance in practice. We look to study resolution techniques empirically in future work.
	The following Section~\ref{sec:sorting} presents the details of the fitness sorting algorithm, which allows such a selection with the time complexity of $\mathcal O(\mathsf{M})$.
	
	\section{Prioritized Sorting Algorithm}
	\label{sec:sorting}
	In this section, we propose the details of the filtering and prioritized sorting algorithm at the application layer. The algorithm finds the relationships between the attributes that define the informativeness to provide a more comprehensive understanding than the linear weights in Eq.~\eqref{eq:object_relevance}.
	
	\subsection{Warm-up Radix Sorting}
	\label{ssec:radix}
	We present a warm-up solution that orders the list $\mathbf{O^v}$ of objects recently received by a vehicle $v$. The radix sorting algorithm, represented by $Sort()$, arranges the orders of the tuples in $\mathbf{O^v}$. Without loss of generality, the algorithm $Sort(a)$ orders the tuples in $\mathbf{O^v}$ in ascending order in the attribute $a$.
	First, we assume the order \textit{Distance (D)} $\bm{>}_\text{{I}}$ \textit{relative Velocity (V)} $\bm{>}_\text{{I}}$ \textit{diRection (R)} $\bm{>}_\text{{I}}$ \textit{Category (C)} of the four attributes in a tuple. Second, we sort the tuples in $\mathbf{O^v}$ by running $Sort(\text{\textit{C}})$, $Sort(\text{\textit{R}})$, $Sort(\text{\textit{V}})$, and $Sort(\text{\textit{D}})$ sequentially. That is, we start sorting from the least significant attribute and move up to the most significant one. 
	%
	We observe that the running time of this solution is about $\mathcal O(4n)$ (asymptotic notation) since the running time of the radix sorting algorithm $Sort()$ is $\mathcal O(n)$ and we need to process four attributes. However, a drawback to this solution is the assumption of a monotonic relationship between the four attributes, i.e., attribute $a$ is more important than another $b$, regardless of the actual value of $b$. 
	
	\subsection{Weighted Fitness Sorting}
	\label{ssec:fitness}
	\pengyuan{To improve the comprehension of the informativeness,} we next introduce a more advanced method, which not only considers the impact of the values of the four attributes but also has a faster running time compared to the radix algorithm. To do so, we first must assume that we have a labeled dataset $\mathbb{D}$ in which a large number of tuples $P^{\star}_{i}$, $i=1,2,...,N$ are collected from vehicles and then labeled as either \emph{Requires Attention} or \emph{Does Not Require Attention} by human experts on traffic safety analysis. If $L^{\star}_i$ denotes the label of tuple $P^{\star}_{i}$, the labeled dataset $\mathbb{D}$ can then be represented as $\mathbb{D} = \{(P^{\star}_{i},L_i^{\star}),i=1,2,..,N \}$. For ease of illustration, we assume $L^{\star}_i\in\{0,1\}$, where $0$ represents Does Not Require Attention and $1$ represents Requires Attention. We note that the dataset $\mathbb{D}$ could be obtained by using the large volumes of traffic data that connected cars stream back to network centers for data analysis~\cite{cheng2018big,sun2018analyzing}. Furthermore, there are numerous semi-supervised classification algorithms~\cite{silva2016survey,triguero2015self} that can be used to build a large dataset $\mathbb{D}$ from a small initial labeled dataset which can be generated manually by traffic experts or even automakers.
	
	Next, we discuss the details of the weighted sorting algorithm. Without loss of generality, we assume that the four attributes \textit{D}, \textit{V}, \textit{R}, and \textit{C} are represented by the $1$st, $2$nd, $3$rd and $4$th attributes, respectively. Given a tuple $P=(D, V, R, C)$ represented by $P=(x_1, x_2, x_3, x_4)$, then a labeled tuple in dataset $\mathbb{D}$ is represented by $(x^{\star}_1, x^{\star}_2, x^{\star}_3, x^{\star}_4,L_i^{\star})$. 
	To better define the relationship between the weights in Eq.~\eqref{eq:object_relevance}, we define a filter $\mathbb{F}$ which weights the four attributes and their relationship by a $4\times 4$ matrix $M$ shown as follows
	\begin{equation}\label{eq:relation_matrix}
		M = \left[ {
			\begin{array}{*{20}{c}}
				m_{1,1}&{m_{1,2}}&{m_{1,3}}&{m_{1,4}}\\
				m_{2,1}&{m_{2,2}}&{m_{2,3}}&{m_{2,4}}\\
				m_{3,1}&{m_{3,2}}&{m_{3,3}}&{m_{3,4}}\\
				m_{4,1}&{m_{4,2}}&{m_{4,3}}&{m_{4,4}}
		\end{array}} \right],
	\end{equation}
	where $m_{i,j}>0$ ($i,j\in\{1,2,3,4\}$) is the weight assigned to the relation between the $i$-th and $j$-th attributes. Given a tuple $P=(x_1, x_2, x_3, x_4)$, filter $\mathbb{F}$ shall compute a fitness $M$ using Mahalanobis distance~\cite{de2000mahalanobis}. The Mahalanobis distance refers to a distance measuring the correlation among the features. It describes the quadratic forms in Gaussian distributions, where the matrix $M$ plays the role of the inverse covariance matrix~\cite{weinberger2009distance}, as follows:
	\begin{flalign}
		&\mathbb{F}(P)= P\times M \times P'\label{eq:fitness_fun}\\
		& =\left[
		{x_1} \; {x_2} \; {x_3} \; {x_4}
		\right]
		\left[ {
			\begin{array}{*{5}{c}}
				m_{1,1}&{m_{1,2}}&{m_{1,3}}&{m_{1,4}}\\
				m_{2,1}&{m_{2,2}}&{m_{2,3}}&{m_{2,4}}\\
				m_{3,1}&{m_{3,2}}&{m_{3,3}}&{m_{3,4}}\\
				m_{4,1}&{m_{4,2}}&{m_{4,3}}&{m_{4,4}}
		\end{array}} \right]
		\left[
		\begin{array}{*{5}{c}}
			x_1\\x_2\\x_3\\x_4
		\end{array}
		\right]\nonumber
	\end{flalign}
	For ease of illustration, we use $d(P_i,P_j)$ to represent the difference of the fitness between two tuples $P_i$ and $P_j$, i.e.,
	\begin{equation}\label{distance_fitness}
		d(P_i,P_j)=(P_i-P_j)\times M \times (P_i-P_j)'
	\end{equation}

	
	\begin{algorithm}[t!]
		\caption{\sysname full-stack filtering algorithm}
		\label{alg:stack}
		\SetKwProg{proc}{thread}{:}{}
		\SetKwProg{main}{Main}{:}{}
		\BlankLine
		\BlankLine
		\nonl {\textbf{Networking layer filtering}}\\
		\proc{CMR}{
			Determine drop or forward\&process \textit{PKT} according to the CMR protocol (see Section~\ref{ssec:routing})\;\label{alg:protocol:lab1} 
		}
		\SetKwProg{proc}{thread}{:}{}
		\SetKwProg{main}{Main}{:}{}
		\BlankLine
		\BlankLine
		\nonl {\textbf{Application layer filtering}}\\
		\proc{Fitness calculation}{
			Update the fitness matrix $M$ (see Eq.~\eqref{eq:relation_matrix}) with new datasets according to Eq.~\eqref{eq:max_problem_for_matrix}, offline\;
		}
		\BlankLine
		\proc{Weighted sorting}{
			Calculate the fitness values of the objects in received messages according to Eq.~\eqref{eq:fitness_fun}\;
			Sort the fitness values\;
		}
		\BlankLine
		\proc{Display}{
			Display the first $L$ objects in the sorted queue to achieve Eq.~\eqref{eq:opt_prob_obj} under the constraint of Eq.~\eqref{utility:constraint:obj_limit}\; 
		}
	\end{algorithm}
	
	We find suitable values for the matrix $M$ used by filter $\mathbb{F}$ by leveraging the knowledge from the labeled dataset $\mathbb{D}$. Let $\mathbb{D}_0$ represent a subset of $\mathbb{D}$ where each tuple is labeled with $0$, and $\mathbb{D}_1=\mathbb{D} \setminus \mathbb{D}_0$. The idea is to find a matrix $M^\star$ that maximizes the difference of the fitness between the subset $\mathbb{D}_0$ and $\mathbb{D}_1$, i.e., the fitness of the tuples in $\mathbb{D}_0$ shall be as small as possible, while that of the tuples in $\mathbb{D}_1$ can be as large as possible, or the vice-versa. More formally, we solve the following optimization problem.
	\begin{align}
		&\mathop {\arg \max}\limits_M \sum\limits_{\substack{i\neq j, L^{\star}_i\neq L^{\star}_j \\ \vb{i \in \mathbb{D}_0, j \in \mathbb{D}_1}}} d(P^{\star}_i,P^{\star}_j) - \sum\limits_{\substack{
				i\neq j, L^{\star}_i = L^{\star}_j \\ \vb{i, j \in \mathbb{D}_0}}} d(P^{\star}_i,P^{\star}_j)
		\label{eq:max_problem_for_matrix}
	\end{align}
	The formula on the left side of the subtraction sign defines the distance between two tuples with different labels (one tuple is labeled $0$ and the other $1$), while the formula on the right side defines the distance between two tuples with the same label (the tuples are labeled either $0$ or $1$). This is a classic metric learning problem for which numerous numerical optimization algorithms are available~\cite{kulis2013metric}. This analysis can be carried out offline, and the obtained $M$ matrix can be used to quickly evaluate whether a newly received tuple $P$ needs further processing by calculating its fitness as shown in Eq.~\eqref{eq:fitness_fun}.
	
	Next, we analyze the time complexity of the weighted sorting solution. As mentioned, obtaining $M$ offline allows us to neglect its computational cost. Upon obtaining $M$, the cost is $\mathcal O(1)$ to compute the fitness value of a tuple, and thus $\mathcal O(N)$ to calculate the fitness values of the $N$ tuples in the $\mathbf{\mathcal O^v}$ list. In addition, the cost to sort the fitness values using the radix sorting algorithm is $\mathcal O(N)$. Hence, the total computational cost is $\mathcal O(2N)$, which is less than that of the warm-up solution (which is $\mathcal O(4N)$). Furthermore, the weighted solution is advantageous as it considers the values in each attribute and their relations. As such, we summarize the full-stack filtering algorithm of \sysname in Algorithm~\ref{alg:stack}.
	
	\section{Evaluation}
	\label{sec:evaluation}
	
	\subsection{Implementation}
	\label{ssec:implementation}
\begin{figure*}[!t]
		\centering
		\begin{subfigure}{0.32\textwidth}
			\includegraphics[width=\linewidth]{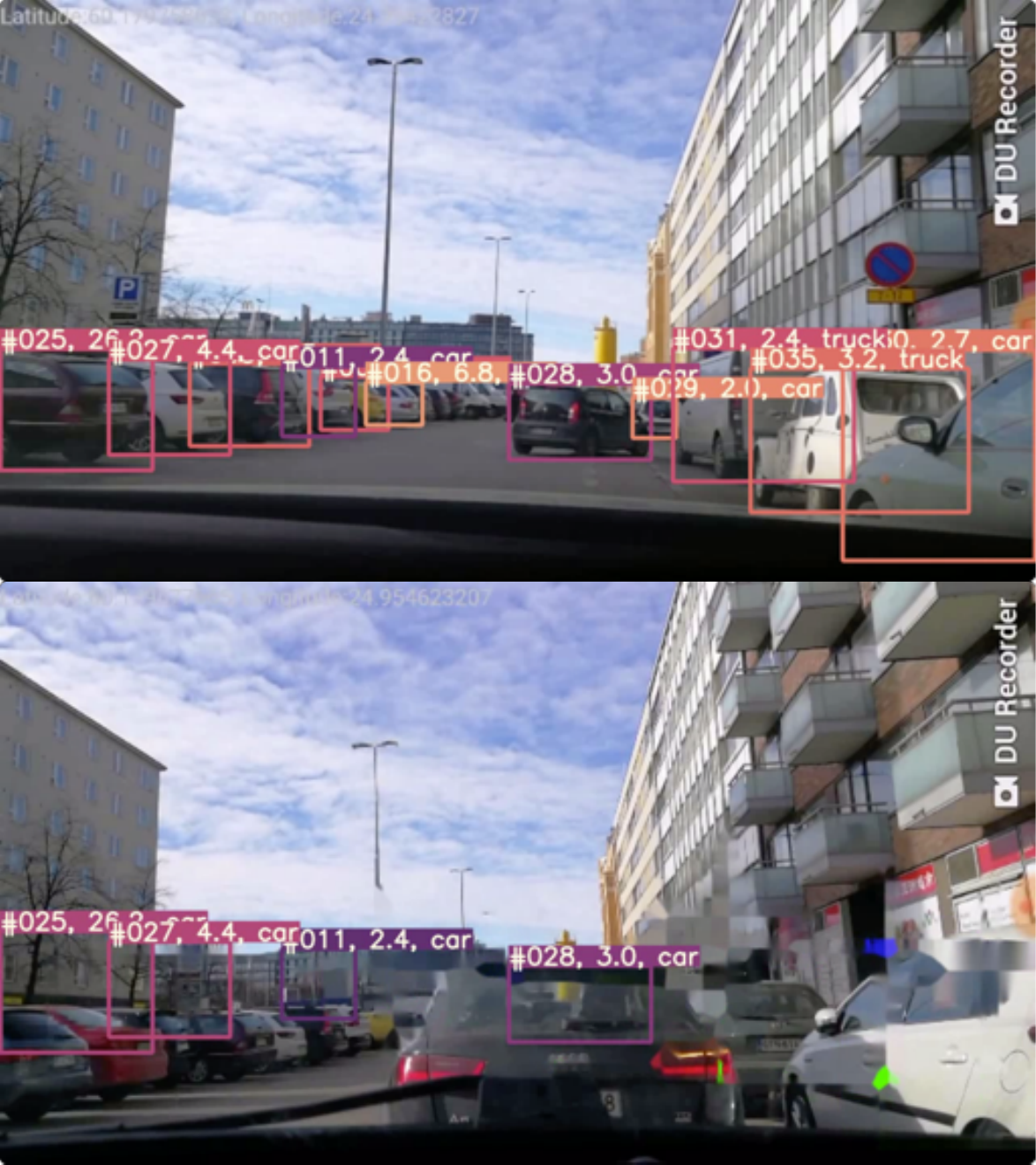}
			\caption{Parking lot.}
			\label{fig:parking}
		\end{subfigure}
		\begin{subfigure}{0.32\textwidth}
			\includegraphics[width=\linewidth]{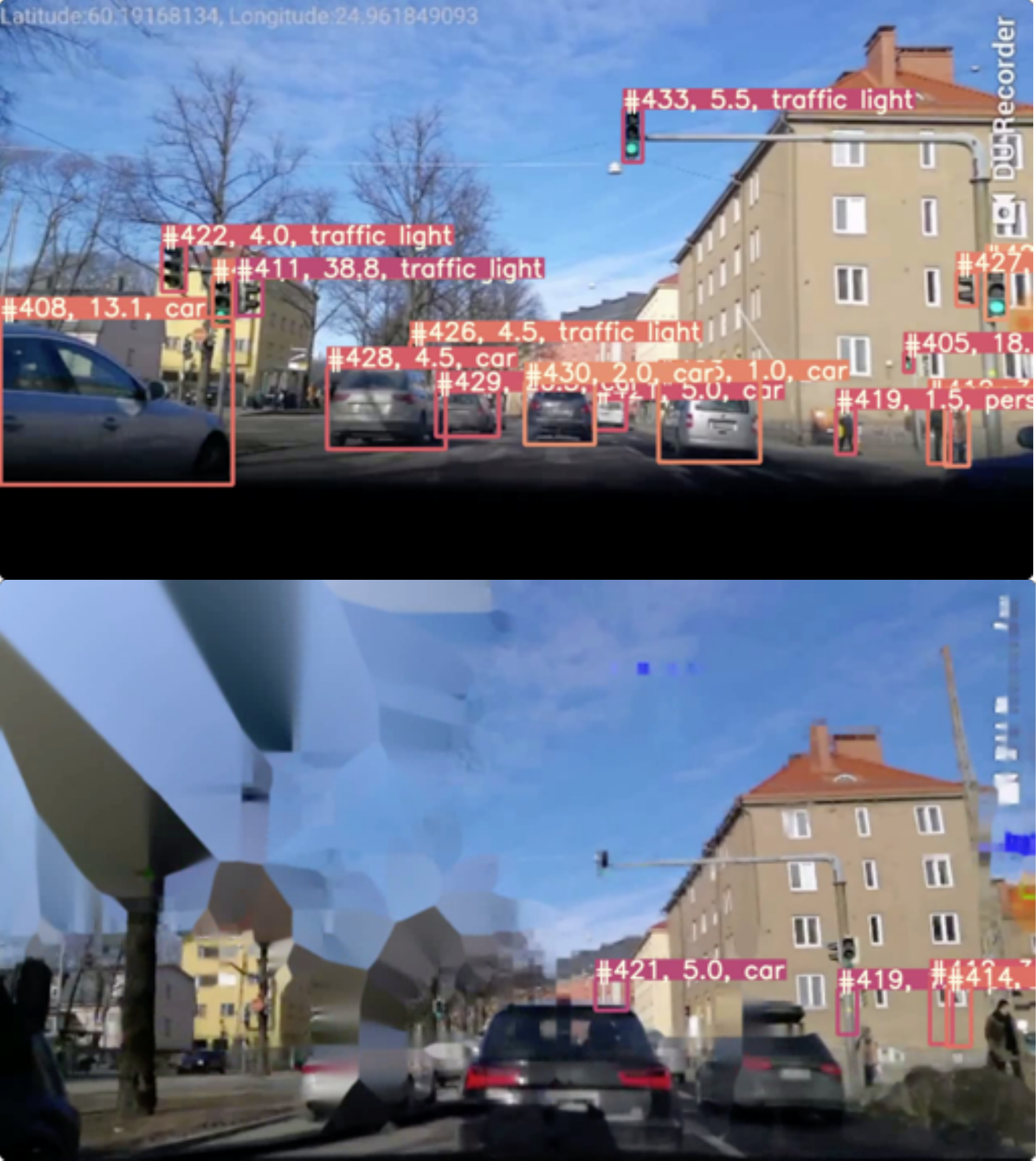}
			\caption{Intersection.}
			\label{fig:intersectoin}
		\end{subfigure}
		\begin{subfigure}{0.32\textwidth}
			\includegraphics[width=\linewidth]{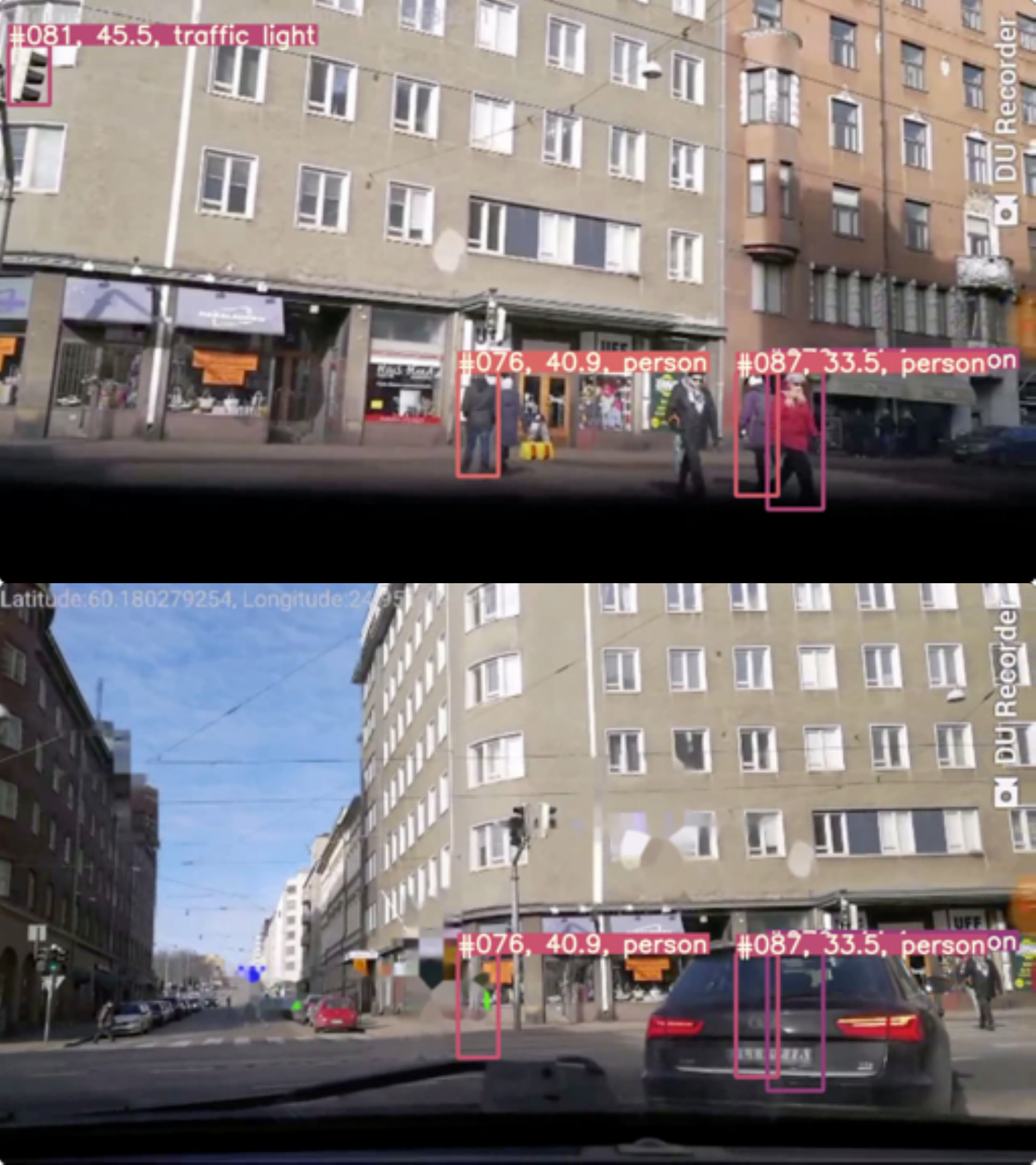}
			\caption{Pedestrians walking across.}
			\label{fig:pedestrian_imp}
		\end{subfigure}
		\caption{POC prototype performance in different contexts. At the top, we see all the objects detected by a \emph{sender} vehicle, whereas at the bottom, a \emph{receiver} vehicle sees only the objects that result from our filtering algorithm. For the sake of comprehensibility, we show all the objects that the sender vehicle detects, although a vehicle does not need to render its own objects.}
		\label{fig:demo}
		\vspace{-4mm}
	\end{figure*}

	Following the system design in Section~\ref{sec:system}, we implement a POC prototype of \sysname as a sender/receiver system. 
	\pengyuan{First, we built a basic prototype that could share vehicular perception in real-time~\cite{9163287}. We deployed the object detector of the sender on a Linux platform with the GPU implementation of Yolo-v5~\cite{glenn_jocher_2020_3983579} to conduct object detection in real-time. We used OpenCV for general image processing such as perspective transformation (from one vehicle to another). The detection outputs tuples that consist of the positions and the labels of the objects, as well as confidence scores. The receiver and the other parts of the sender (camera, IMU data collector) were implemented on the Android platform to simulate the hardware and software environment of the vehicular equipment for augmented vision. We used the GPS sensor to report the GPS coordinates of the vehicle, and the monocular camera to capture the front-facing view from the vehicle. We used OpenGL to render the augmented information on top of the camera view. Our implementation operates with a monocular camera (on the Android phone), while vehicles with stereoscopic cameras could make transformations easier. We drove two vehicles equipped with a sender and a receiver respectively across Helsinki city center with one vehicle following the other. The sender detected objects in its line of sight and shared them with the receiver in real-time. We recorded the video including all shared objects without AICP as a benchmark (as shown in the top of Figure~\ref{fig:demo})\footnote{\pengyuan{As our previous work demonstrated~\cite{9163287}, the overall delay of V2V perception sharing, including image capturing, processing, data transmission, and rendering, is 57.7 ms.}\label{footnote:577}}. Next, we modified the architectures of the sender and receiver to realize the full AICP system as follows.}
	
\begin{figure}[!t]
		\centering
		\includegraphics[width=.7\linewidth]{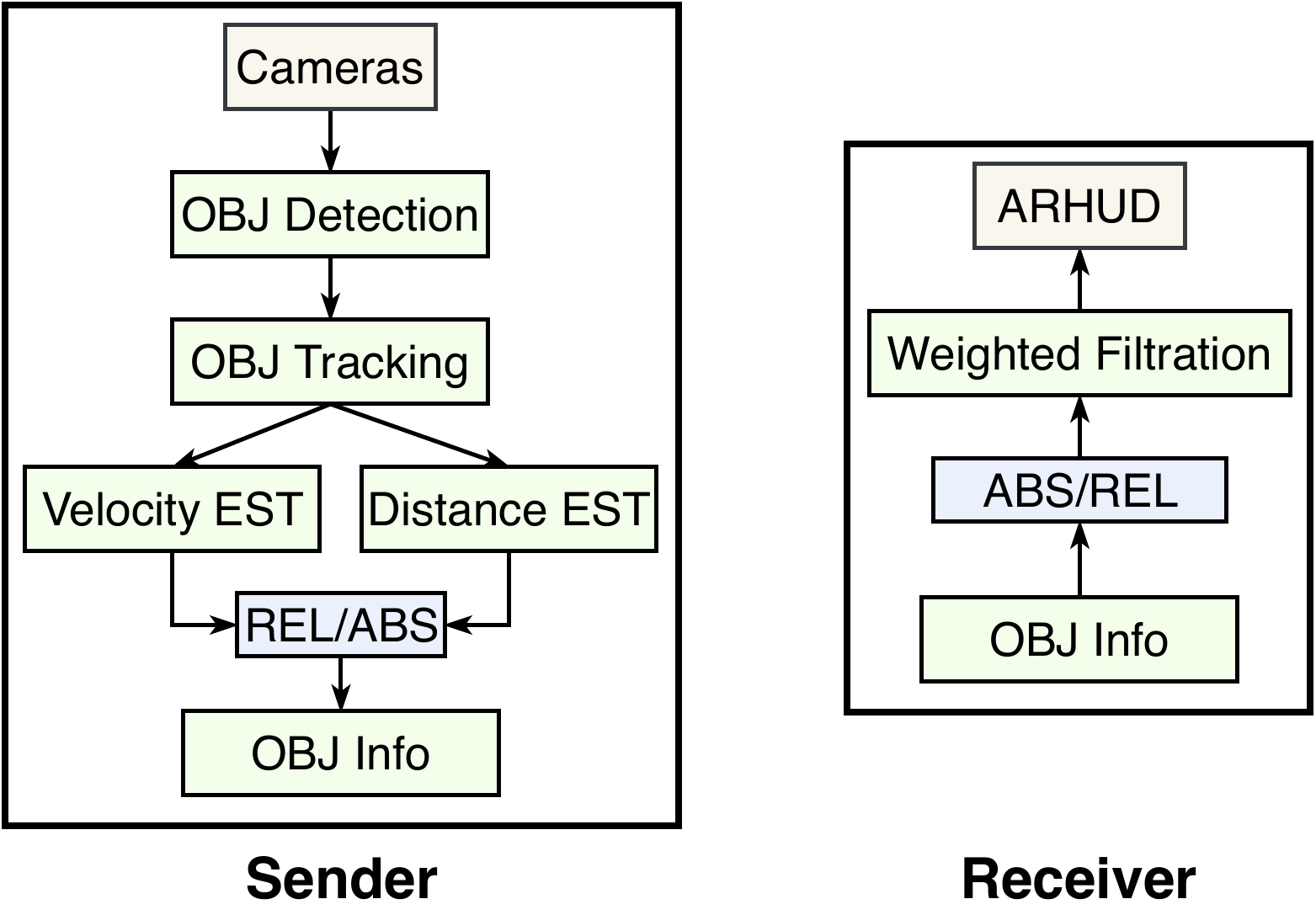}
		\caption{Data flows of the POC prototype system including the practical data processing steps such as object (OBJ) detection, velocity estimation (EST), and relative to absolute value transformation (REL/ABS).}
		\label{fig:poc-data-flow}
	\end{figure}

	%
	\fakeparagraph{Sender.}
	As shown in Figure~\ref{fig:poc-data-flow}, the sender consists of five components: 
	\begin{enumerate*}[label=(\roman*)]
		\item object detection, 
		\item object tracking,
		\item distance estimation,
		\item velocity estimation, and
		\item relative/absolute value transform
	\end{enumerate*}.
	The system feeds the results to the \emph{object tracking} component, which maintains a list of the objects tracked in the previous frames. First, such a component calculates the Intersection over Union~(IoU), defined as $\frac{\text{area of overlap}}{\text{area of union}}$, between the objects in the previous and the latest frames. Next, it uses a greedy algorithm (quicksort) to sort the similarity according to the IoU scores~\cite{bochinski2017high}.
	The system passes the tracking result to the \emph{velocity estimation} and \emph{distance estimation} components simultaneously. The velocity estimation uses the historical positions of an object to estimate the speed and direction of its movement. The distance estimation uses the object's position and the ratio between the object area and the average size of the object to estimate the distance to the object. \pengyuan{As a showcase,} we categorize the distance into three classes, namely nearby, middle, and far away. After gathering the objects' positions, labels, velocities, and distances, the sender broadcasts the messages. 
	
	\fakeparagraph{Receiver.}
	The receiver consists of 
	\begin{enumerate*}[label=(\roman*)]
		\item value transform, 
		\item weighted filtration, and 
		\item AR~display.
	\end{enumerate*}
	The receiver applies the \textit{weighted fitness sorting} algorithm to select the top $L$ informative objects out of the received message and renders them to the driver. 
	
	\fakeparagraph{Dataset.}
	\benjamin{To show the performance of the \textit{weighted fitness sorting} algorithm~(Section~\ref{ssec:fitness}), we require that the dataset is labeled. However, due to the difficulty of labeling such data without safety experts, we instead use a heuristic approach just to illustrate the potential performance. Specifically, we label an object as ``require attention'' when it has a distance of less than 23 meters (safe breaking distance when driving at a safe speed~13 m/s), or velocity larger than 13 m/s, or is a pedestrian and ``not require attention'' in other cases.} As a result, we get 22996 and 52880 objects for the two categories, respectively. \benjamin{We note that the real world distribution might be different but we believe the heuristic provides a reasonable approximation for our purposes.}
	
	\fakeparagraph{Algorithm.}
	We use the \texttt{metric-learn} library contributed by the authors of~\cite{de2020metric} to implement LMNN for computing the Mahalanobis distance matrix (see Eq. (8) to Eq. (10)). The matrix computation takes 225.79 seconds per 10000 objects. As described in Section~\ref{ssec:fitness}, the matrix computation is offline, and thus has no impact on the real-time system performance.
	Using the learned matrix, each receiver can sort 100 received objects within a millisecond on average. We also open source the code of the weighted fitness sorting algorithm and the 75876-object data extracted from the video\footnote{\label{fn:aicp_code}\url{https://github.com/pengyuan-zhou/AICP}}.
	
	\subsection{Showcase Performance}
	\label{ssec:showcase}
	
	Table~\ref{tab:poc-latency} summarizes the latencies of the processes. As shown, the overall latency in the sender system is only $12.6$ milliseconds and thus has a negligible impact on information dissemination\footnote{\pengyuan{Together with the original V2V perception sharing delay, 57.7 ms~(\ref{footnote:577}), the overall delay is only 70.3 ms which is much smaller than the common standard of V2V safety applications (100 ms).}}. The additional latency added at the receiver end is mainly the sort latency, which is $\sim1$ millisecond on average and does not affect the networking performance.
	
	As shown in Figure~\ref{fig:demo}, \sysname effectively simplifies and improves the comprehensibility of the cooperative perception system by pruning shared perception information and showing only the most critical objects. In different contexts -- e.g., in a parking lot, at an intersection, and with pedestrians crossing the street -- the filtered objects that are displayed are much easier to comprehend, and thus facilitate driving. In comparison, cooperative perception systems without~\sysname would display numerous objects as in Figure~\ref{fig:parking} and Figure~\ref{fig:intersectoin}, or uninformative objects such as the traffic lights in Figure~\ref{fig:pedestrian_imp}. Note that in different areas the system can improve performance by \pengyuan{easily} adapting $L$ (the number of selected objects defined in Eq.~\eqref{utility:constraint:obj_limit}). For instance, $L$ might be smaller in a parking lot with fewer moving objects in comparison with an intersection that will likely have more moving objects. Cloud services like Google Map can be used for area identification.
	
	\begin{table}[!t]
		\renewcommand*{\arraystretch}{1.2}
		\caption{Latency breakdown of the POC steps.}
		\centering
		\begin{tabular}{|c|c|c|}
			\specialrule{1.3pt}{1pt}{1pt}
			&\textbf{Task}    & \textbf{Execution Time (ms)} \\ \hline
			\multirow{ 4}{*}{Sender}
			&Pre-processing                    & $1.1$                  \\ \cline{2-3}
			&Object Detection                 & $6.9$                  \\ \cline{2-3}
			&Object Tracking                  & $2.4$                  \\ \cline{2-3}
			&Velocity \& Distance Est.        & $1.2$    \\
			\hline
			\multirow{ 1}{*}{Receiver} 
			&Sorting & $1$ \\
			\hline \hline
			\multirow{ 1}{*}{\sysname} 
			&Overall Latency                  & $12.6$                 \\
			\specialrule{1.3pt}{1pt}{1pt}
		\end{tabular}
		\label{tab:poc-latency}
	\end{table}
	
	\benjamin{Unfortunately, comparing AICP with baselines from literature is difficult because existing works in cooperative perception (see the related works section) typically have related but not matching goals. For example, Aoki et al.~\cite{aoki2020} aim to reduce V2V network congestion by using deep learning to stop redundant (same object) message transmission. In contrast, we focus on the complementary goal of preventing uninformative message transmission (and therefore, we have different metrics). Additionally, performance comparisons with literature baselines or a purpose-built baseline (such as a deep learning model) would not be overly informative since, as previously mentioned, we do not have a real-world dataset with empirical informativeness labels or rankings. Thus comparisons would not be readily generalizable.}
	
	\subsection{Simulation}
	\label{subsec:simulation}
	Following the performance of the prototype shown in Section~\ref{ssec:showcase}, we next illustrate the performance of \benjamin {CMR} through a larger scale simulation. We open-source the core scripts and datasets of the simulation\footref{fn:aicp_code}. 
	
	\fakeparagraph{Simulation Setup.}
	As pointed out in Section~\ref{ssec:routing}, CMR focuses on filtering low-informativeness packets rather than improving communication efficiency. To isolate the effect of CMR, we exclude any communication efficiency-focused routing protocols such as GPSR~\cite{karp2000gpsr} and DV-CAST~\cite{tonguz2010dv}. Instead, we compare the transmission statistics with and without CMR in city-scale V2V simulations. We select a $4x5$ km rectangular area of London city center and generate traffic utilizing a real-world dataset\footnote{\url{https://data.gov.uk/dataset/gb-road-traffic-counts}}. 
	
	\fakeparagraph{Analogue models} We simulate the traffic during an off-peak period  ($7$ am) and a peak period ($6$ pm) using \texttt{Veins}~\cite{sommer2011bidirectionally}, an open-source framework for running vehicular network simulations. \texttt{Veins} is based on \texttt{OMNeTpp}, an event-based network simulator, and \texttt{SUMO}, a road traffic simulator. We simulate for $60$ seconds. To ensure realism, we employ the two-ray interference model~\cite{sommer2011using} for radio propagation. The model improves over the vanilla two-ray ground model by  capturing the ground reflection effects. We also employ the obstacle shadowing model~\cite{sommer2013ivc} to capture the effects of buildings on signal transmissions. The upper part of Table~\ref{tab:parameter} details the parameters.
	
	\fakeparagraph{Routing protocol} We use IEEE 802.11p as the base networking protocol for V2V communications. Following the design of CMR, we set the hop limit of each message to $2$, the maximum concerned source distance to $100$ meters, and the maximum heading direction difference between the source vehicle and the receiver to $30$ degrees.
	The C-V2X standard~\cite{gsma-cv2x,etsi-its-g5-5g} recommends a message frequency between 1 and 10 Hz. To test the baseline performance, we let each vehicle broadcast VDUs at 10 Hz. From the POC test, we observe 10 objects per image on average. Thus, the average VDU packet size is set to 102 bytes. Table~\ref{tab:parameter} and Table~\ref{tab:pkt} detail the CMR parameters and packet format.
	
	\begin{table}[!t]
		\renewcommand*{\arraystretch}{1}
		\caption{Simulation parameters}
		\label{tab:parameter}
		\centering
		\begin{tabularx}{8cm}{|l|X|}
			\specialrule{1.3pt}{1pt}{1pt}
			\textbf{Simulation Parameter} & \textbf{Value}  \\
			\hline
			Radio Propagation Model & Two-Ray Interference Model~\cite{sommer2011using} \\
			Shadowing Model & Obstacle Shadowing Model~\cite{sommer2013ivc} \\
			IEEE 802.11p Bit Rate  & $6$ Mbps     \\   
			Transmission Power    & $20$ mW       \\
			Noise Floor     & $-98$ dBm           \\
			Antenna Height   & $1.5$ m       \\
			Antenna Type & Front-Rear\footref{fn:aicp_code} \\ 
			Number of Vehicles & 61 (7am), 212 (6pm)\footref{fn:aicp_code} \\ ~\cite{code} \\
			Simulation Area  & $4000$x$5000$ m        \\
			Simulation Time  & $60$ s \\
			\hline
		\end{tabularx}\par\vskip-1.4pt
		\begin{tabularx}{8cm}{|l|X|}
			\specialrule{1.3pt}{1pt}{1pt}
			\textbf{CMR Parameter} & \textbf{Value} \\ \hline
			Beacon Generation Rate& $10$ Hz\\
			Hop Limit & $2$ \\
			Max Source Heading Direction Deviation & $30$°   \\   
			Max Source Distance    & $100$ m      \\
			Packet Size & $102$ Bytes  \\
			\hline
		\end{tabularx}
	\end{table}
	
	\begin{table}[!t]
		\renewcommand*{\arraystretch}{1}
		\caption{Packet format ($102$ B)} 
		\label{tab:pkt}
		\centering
		\begin{tabular}{|L{1.1cm}|C{3cm}|C{2.6cm}|}
			\specialrule{1.3pt}{1pt}{1pt}
			\textbf{Field}  & \textbf{Metadata}  & \textbf{Size (Byte)}
			\\ \hline
			\multirow{ 7}{*}{Object}
			& ID             & $2$        
			\\ \cline{2-3}
			&position\_x     & $1$        
			\\ \cline{2-3}
			&position\_y     & $1$    
			\\ \cline{2-3}
			&velocity        & $1$      
			\\ \cline{2-3}
			&distance        & $1$       
			\\ \cline{2-3}
			&label           & $1$             
			\\ \cline{2-3}
			&confidence      & $1$      
			\\ \hline
			\multirow{3}{*}{Vehicle}
			&IMUs           & $12$       
			\\ \cline{2-3}
			& Timestamp      & $2$  
			\\ \cline{2-3}
			&GPS             & $8$      \\
			\hline
		\end{tabular}
	\end{table}
	
	\begin{figure}[!t]
		\centering
		\includegraphics[width=.4\linewidth]{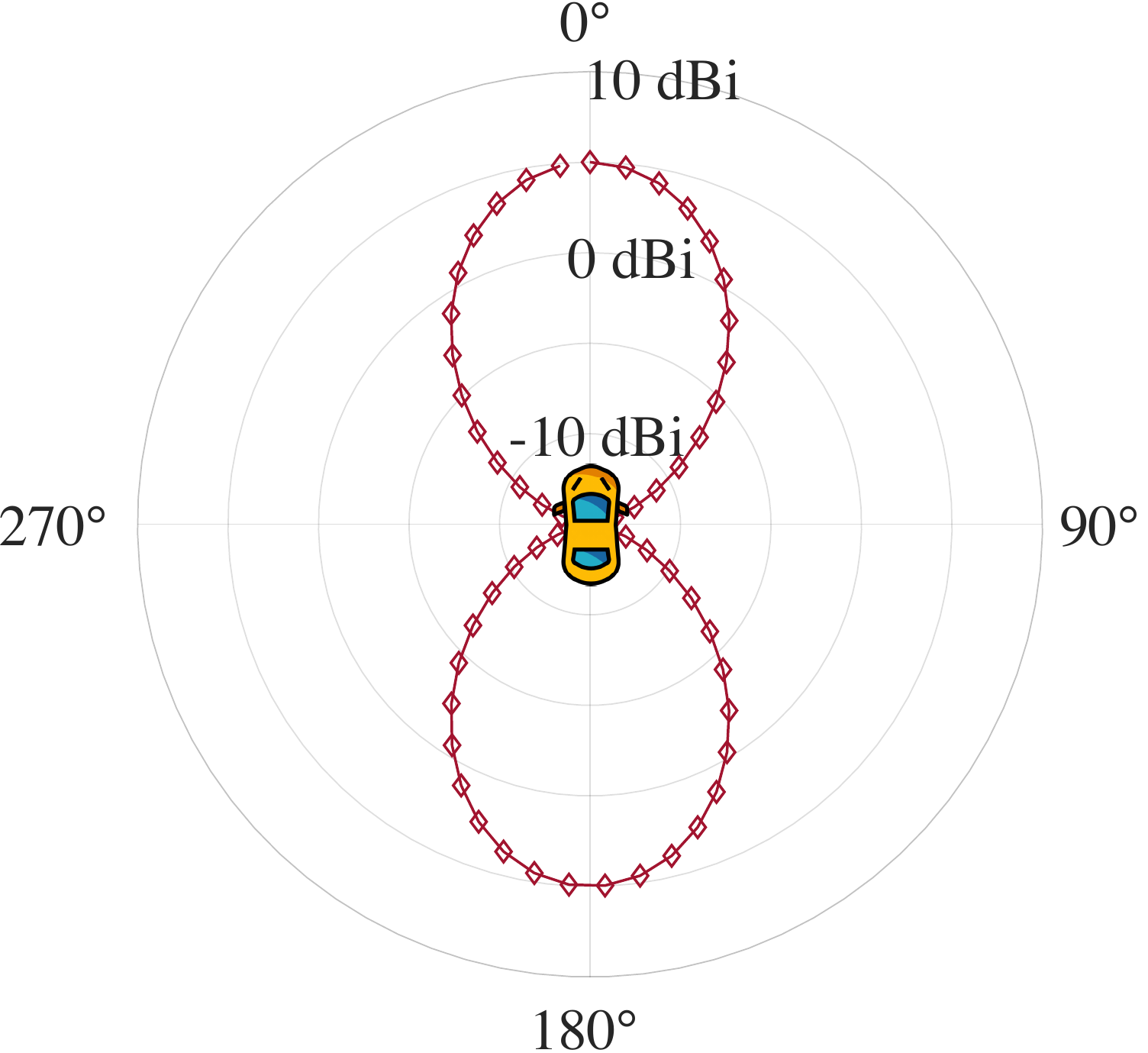}
		\caption{Front-Rear antennae radiation pattern (view from above with vehicle facing north) in dBi.}
		\label{fig:antenna}
	\end{figure}
	
	\fakeparagraph{Results}
	Due to the dense traffic in the selected area, the average vehicle speed in the simulation period is about $6$ km/h, thus reflecting the lower bound of system performance under extremely congested scenarios.
	
	\fakeparagraph{Antenna type} As demonstrated by~\cite{kornek2010effects,eckhoff2016impact}, angled antennae, compared to idealistic isotropic antennas, can significantly change the vehicular network dynamics. Therefore, we propose to use a Front-Rear\footref{fn:aicp_code} antenna to complement the CMR protocol. 
	Recall that CMR prioritizes data sent by source vehicles traveling in a similar direction as the receiver. As shown in Figure~\ref{fig:antenna}, Front-Rear amplifies the signal in the forward and rear directions while reducing the transmission range on the sides. Hence, Front-Rear reduces the packets sent from vehicles driving on the sides and reduces the burden of the filters. Front-Rear can be deployed similarly as Patch~\cite{kornek2010effects}, i.e., mounted to the front of the right and left side mirrors and the right and left side of the rear windshield.
	
	\begin{figure}[!t]
		\centering
		\begin{subfigure}{0.46\textwidth}
			\includegraphics[width=\linewidth]{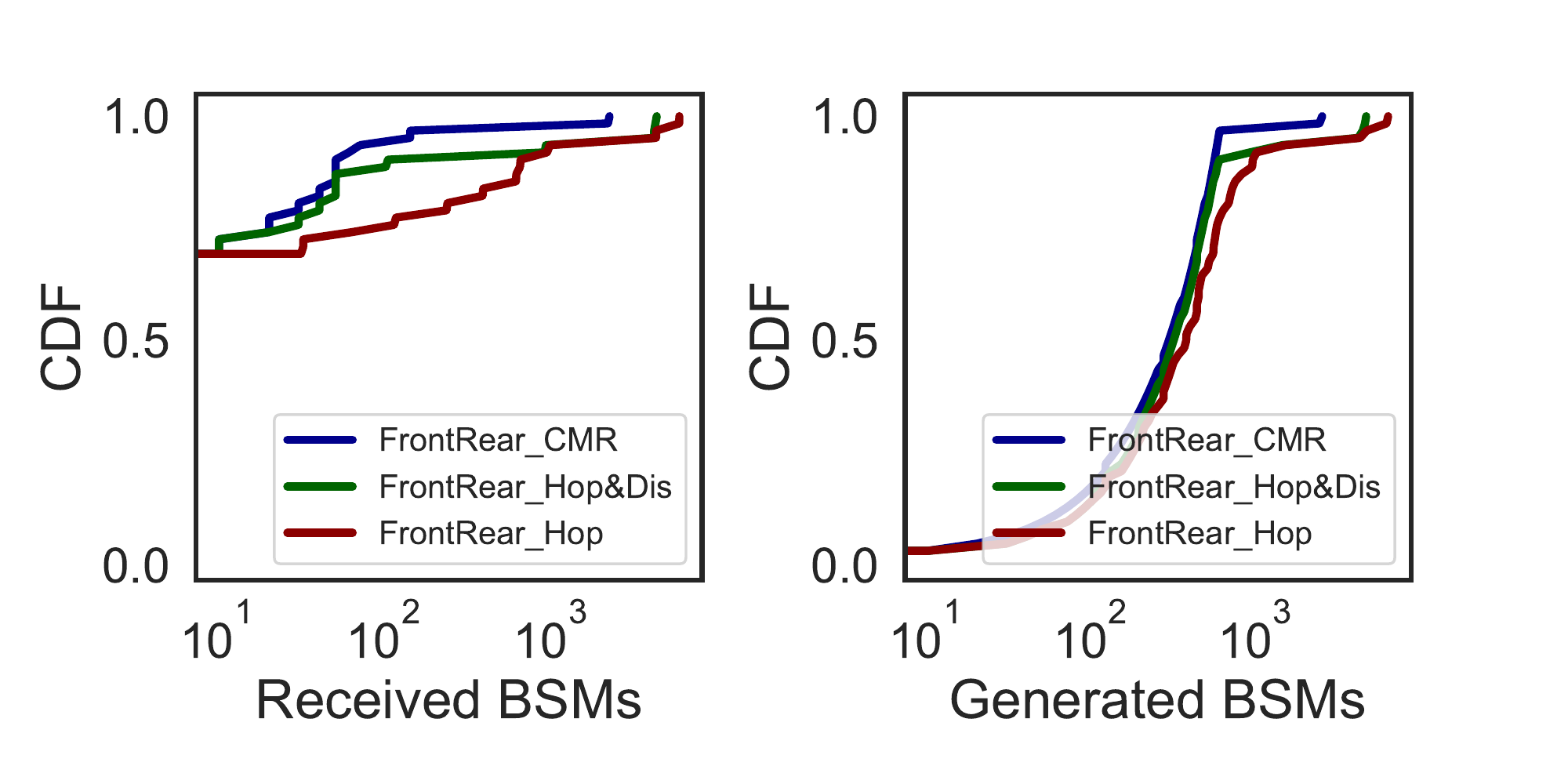}
			\caption{}
			\label{sfig:bsm_7am}
		\end{subfigure}
		\begin{subfigure}{0.46\textwidth}
			\includegraphics[width=\linewidth]{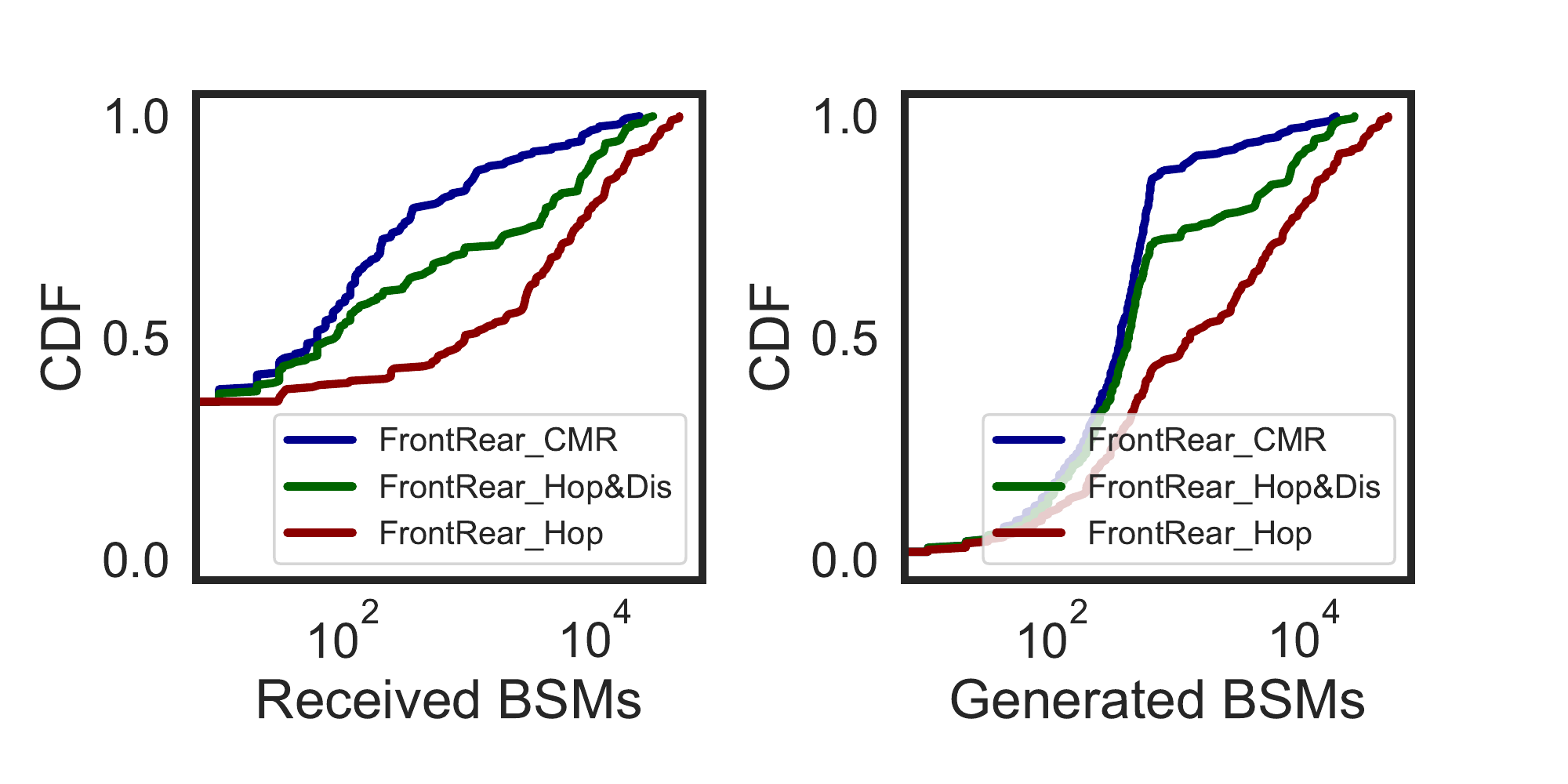}
			\caption{}
			\label{sfig:bsm_6pm}
		\end{subfigure}
		\caption{CDF of received and generated basic safety messages using a Front-Rear antenna and different filter mechanisms: the proposed CMR, Hop\&Dis filter with 2 hops and 100 m distance, and Hop filter with 2 hops at \subref{sfig:bsm_7am} 7 am and \subref{sfig:bsm_6pm} 6 pm.}
		\label{fig:bsm}
	\end{figure}
	
	\begin{figure}[!t]
		\centering
		\begin{subfigure}{0.24\textwidth}
			\includegraphics[width=\linewidth]{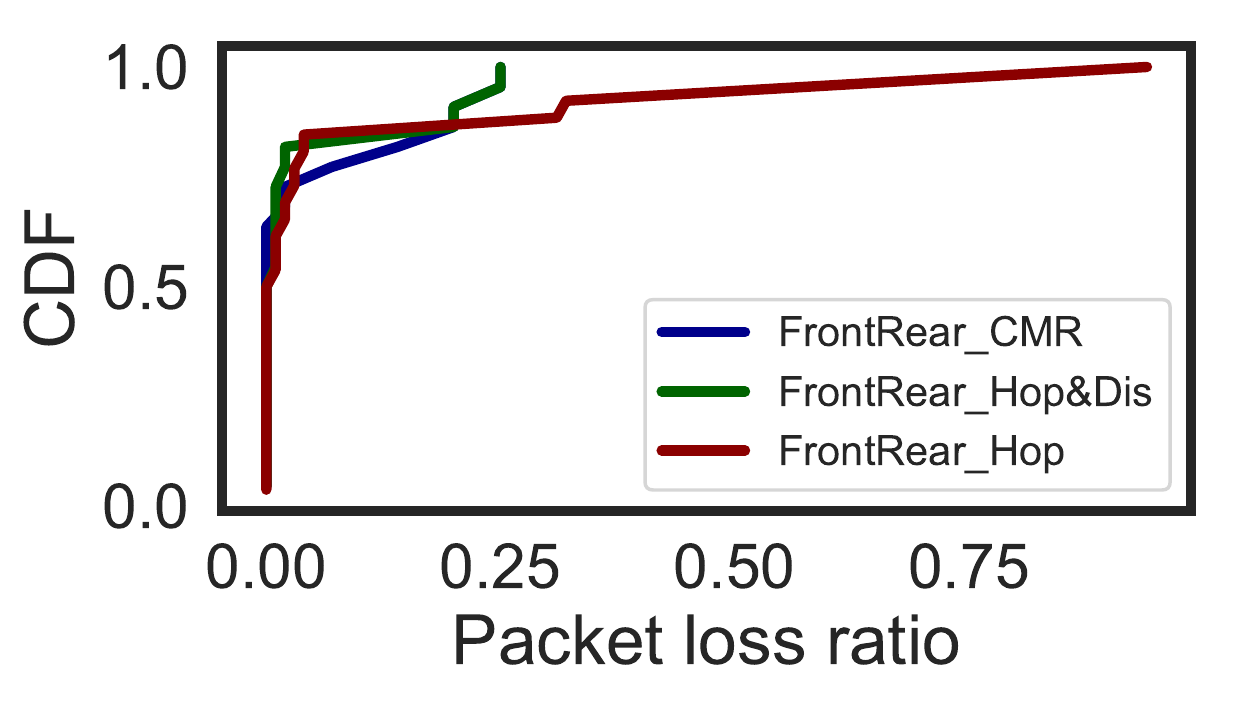}
			\caption{}
			\label{sfig:loss_7am}
		\end{subfigure}
		\begin{subfigure}{0.24\textwidth}
			\includegraphics[width=\linewidth]{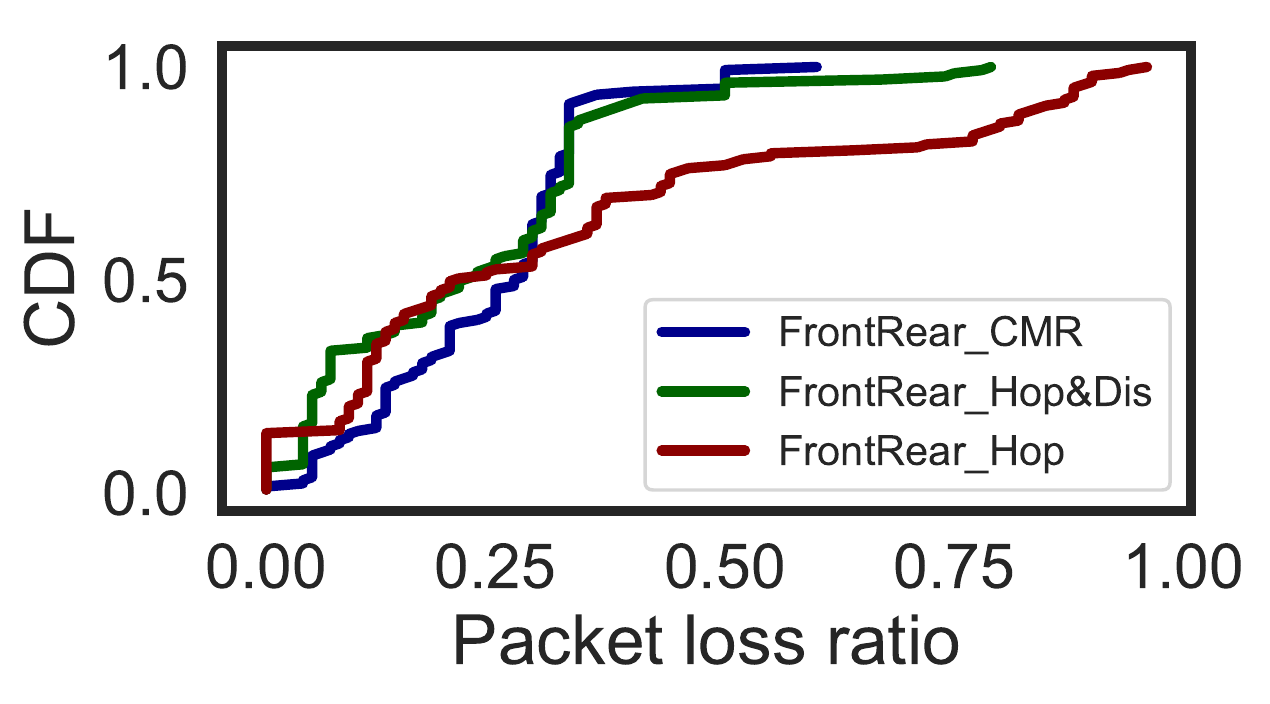}
			\caption{}
			\label{sfig:loss_6pm}
		\end{subfigure}
		\caption{CDF of packet loss ratio with different filters: 
			the proposed CMR, Hop\&Dis filter with 2 hops and 100 m distance, and Hop filter with 2 hops at \subref{sfig:loss_7am} 7 am and \subref{sfig:loss_6pm} 6~pm.}
		\label{fig:loss}
	\end{figure}
	
	\begin{figure}[!t]
		\centering
		\begin{subfigure}{0.24\textwidth}
			\includegraphics[width=\linewidth]{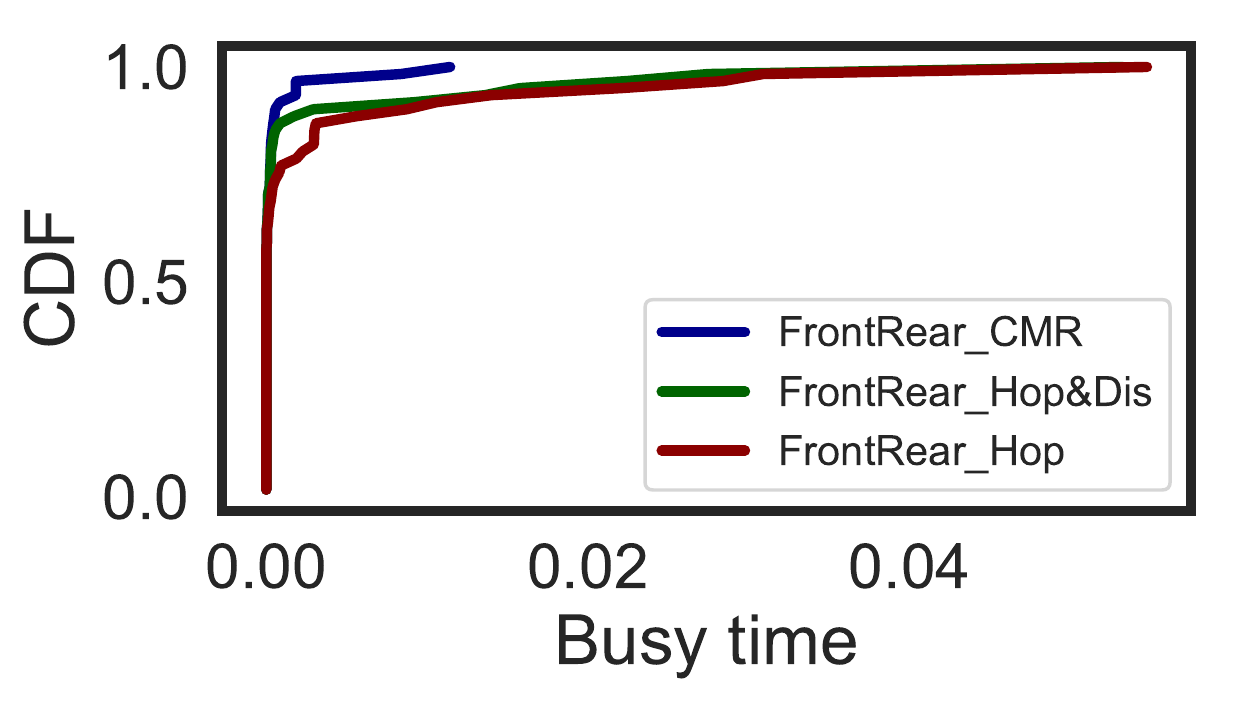}
			\caption{}
			\label{sfig:busy_7am}
		\end{subfigure}
		\begin{subfigure}{0.24\textwidth}
			\includegraphics[width=\linewidth]{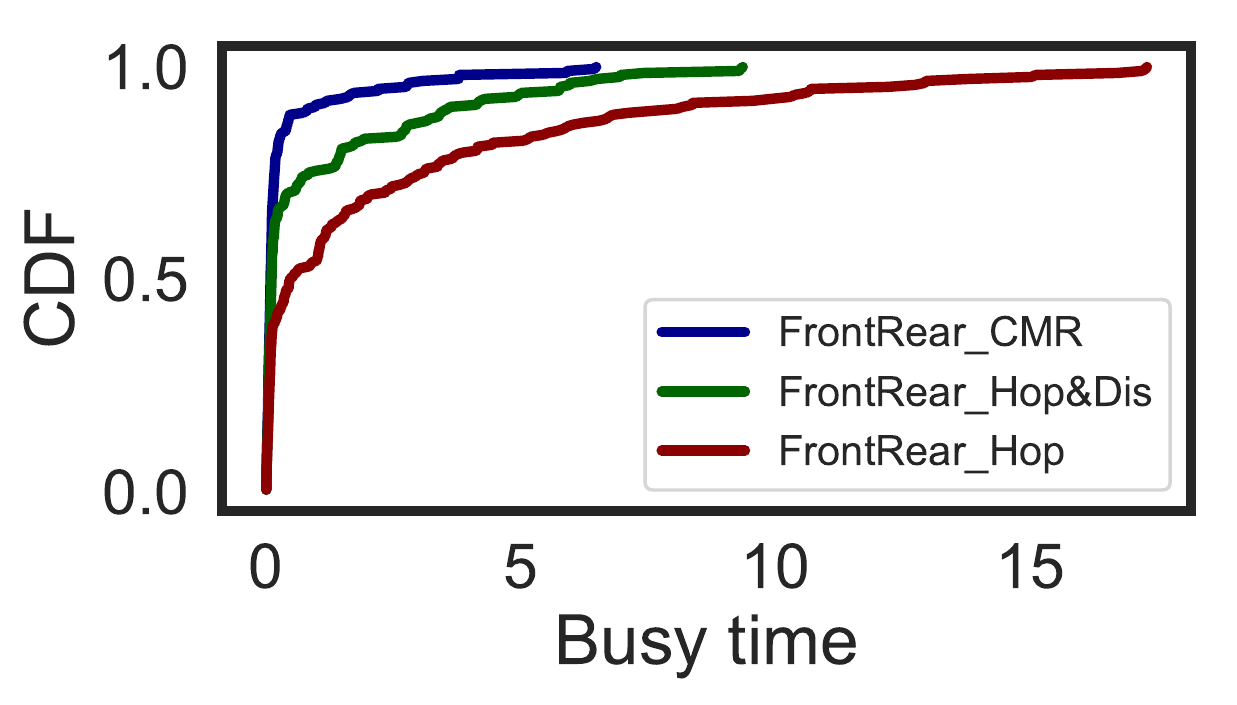}
			\caption{}
			\label{sfig:busy_6pm}
		\end{subfigure}
		\caption{CDF of channel busy time with different filters: 
			the proposed CMR, Hop\&Dis filter with 2 hops and 100 m distance, and Hop filter with 2 hops at \subref{sfig:busy_7am} 7 am and \subref{sfig:busy_6pm} 6~pm.}
		\label{fig:busy}
	\end{figure}
	
	\fakeparagraph{Filters} We compare CMR with two other classic filters: hop and distance limit (Hop\&Dis), and hop limit only (Hop). Figure~\ref{fig:bsm} shows the empirical cumulative distributions of the number of generated and received basic safety messages~(BSMs) by vehicles when using different filters. Figure~\ref{fig:loss} and Figure~\ref{fig:busy} show the empirical cumulative distributions of the packet loss ratio and channel busy time experienced by the vehicles. As summarized in Table~\ref{tab:results}, CMR effectively filters considerably more packets in both off-peak and peak periods compared to Hop\&Dis (-72\% / -61\%) or only Hop (-80\% / -81\%). Furthermore, CMR has a slightly higher packet loss ratio than Hop\&Dis but shows considerably less channel busy time than the other two filters in the peak period (-61\% and -83\%, respectively).
	\begin{table}[!t]
		\renewcommand*{\arraystretch}{1.2}
		\caption{Simulation results (7 am / 6 pm).}
		\centering
		\begin{tabular}{|c|c|c|c|}
			\specialrule{1.3pt}{1pt}{1pt}
			\textbf{Filter}  & \textbf{Received BSMs} & \textbf{Busy time (s)}  & \textbf{Packet loss } \\ \hline
			CMR  & 83 / 1024 & 0 / 0.39 &  0.31 / 0.63 \\
			Hop\&Dis  & 303 / 2630 &  0 / 1.01 &  0.30 / 0.54 \\
			Hop  & 416 / 5652 &  0 / 2.41 & 0.09 / 0.71 \\
			\specialrule{1.3pt}{1pt}{1pt}
		\end{tabular}
		\label{tab:results}
		\vspace{-4mm}
	\end{table}

	As mentioned in Section~\ref{sec:system}, \sysname focuses on informativeness and thus employs CMR to filter low-informativeness packets. We implement CMR in \texttt{Veins} with only five lines of code and argue the CMR would also be lightweight in reality. Hence, \textit{CMR could be easily integrated into routing protocols focusing on communication efficiency improvements}.
	
	\section{Discussion}
	\label{sec:discussion}

	Due to the stochastic nature of human driving and the driving environment, packets containing information about a certain object may not reach their destination consistently. For instance, vehicles may move in and out of the transmission region (hop limit), and thus only receive a fraction of the packets concerning a given object. Similarly, the filtering algorithm may select different objects to display on each round. As such, the objects displayed on-screen may \emph{flicker} and thus significantly degrade the driving experience by distracting the driver and deteriorating the received information quality~\cite{colley2017design}. 
	
	As this contradicts the goals of our proposed \sysname, in future work we look to explore potential solutions such as an object persistence delay. In other words, once an object is displayed on-screen, the object remains displayed for a fixed amount of time, regardless of updates and potential filtering. This delay should be set to a value high enough in order not to distract the driver with high frequency flickering. However, longer delays may lead to a cluttering of the display with objects of low informativeness. A delay between 500\,ms to 2\,s could represent an acceptable trade-off to preserve high informativeness while avoiding flickering.
	
	Beyond the object flickering, other user-centric and \vb{human-computer interaction (HCI)} aspects of the \sysname system or potential system extensions could also be a target of future work. In particular, we could consider two different aspects. First, multi-modal cues (visual, audio, and tactile~\cite{correia2017avui,lateral-position-5444759}) could ease the driver's cognitive load and improve driving performance when the driver's attention primarily focuses on the road~\cite{onimaru2008cross}. Second, we could evaluate the placement of certain visual contents (e.g., focal vs. peripheral placement~\cite{hauslschmid2015augmenting, chaturvedi2019peripheral}) to determine the optimal positioning for driving performance.
	
	Finally, our system focuses on fast information filtering and it relies on reasonable heuristics (e.g., objects physically closer to the vehicle are more informative) to determine the informativeness of any specific object near the vehicle in near real-time. However, given that a group of vehicles is an interacting set of agents, other methods might help predict object informativeness in more complex scenarios such as, for instance, an accident caused by a time series of events. Therefore, in future work, we will examine a data-based deep learning approach that accounts for such complex scenarios. \benjamin{Though importantly, compared to such an approach, AIPC will still likely have some practical advantages such as simplicity, explainability, and lower latency.}
	
	
	
	\section{Conclusion}
	\label{sec:conclusion}
	
	In this work, we propose~\sysname, the first solution that focuses on optimizing informativeness for pervasive cooperative perception systems with efficient filtering at both the network and application layers. To facilitate system networking, we also propose a networking protocol stack that includes VDU and CMR, a dedicated data structure and a lightweight routing protocol, respectively, both designed specifically for informativeness-focused applications. We also formulate the informativeness problem in cooperative perception systems from several different levels and propose a prioritized sorting algorithm for fast information-based filtering. 
	Overall, \sysname displays only the most important information shared by nearby vehicles, and thus eases the understanding of the surrounding for the drivers.
	We implement a POC of our proposal with ARHUD and show that the system has negligible additional processing latency ($12.6$ ms). Additionally, simulation results show that CMR effectively filters less relevant packets, and thus considerably improves the channel availability of the vehicles.


\begin{thebibliography}{10}
\providecommand{\url}[1]{#1}
\csname url@samestyle\endcsname
\providecommand{\newblock}{\relax}
\providecommand{\bibinfo}[2]{#2}
\providecommand{\BIBentrySTDinterwordspacing}{\spaceskip=0pt\relax}
\providecommand{\BIBentryALTinterwordstretchfactor}{4}
\providecommand{\BIBentryALTinterwordspacing}{\spaceskip=\fontdimen2\font plus
\BIBentryALTinterwordstretchfactor\fontdimen3\font minus
  \fontdimen4\font\relax}
\providecommand{\BIBforeignlanguage}[2]{{%
\expandafter\ifx\csname l@#1\endcsname\relax
\typeout{** WARNING: IEEEtran.bst: No hyphenation pattern has been}%
\typeout{** loaded for the language `#1'. Using the pattern for}%
\typeout{** the default language instead.}%
\else
\language=\csname l@#1\endcsname
\fi
#2}}
\providecommand{\BIBdecl}{\relax}
\BIBdecl

\bibitem{8959552}
M.~P. Muresan and S.~Nedevschi, ``Multi-object tracking of 3d cuboids using
  aggregated features,'' in \emph{2019 IEEE 15th International Conference on
  Intelligent Computer Communication and Processing (ICCP)}, 2019.

\bibitem{7995787}
H.~Kim, J.~Cho, D.~Kim, and K.~Huh, ``Intervention minimized semi-autonomous
  control using decoupled model predictive control,'' in \emph{2017 IEEE
  Intelligent Vehicles Symposium (IV)}, 2017, pp. 618--623.

\bibitem{8620918}
A.~Ar\i{}kan, A.~Kayaduman, S.~Polat, Y.~\c{S}im\c{s}ek, I.~C. Dikmen, H.~G.
  Bak\i{}r, T.~Karada\u{g}, and T.~Abbasov, ``Control method simulation and
  application for autonomous vehicles,'' in \emph{2018 International Conference
  on Artificial Intelligence and Data Processing (IDAP)}, 2018, pp. 1--4.

\bibitem{rauch2012car2x}
A.~Rauch, F.~Klanner, R.~Rasshofer, and K.~Dietmayer, ``Car2x-based perception
  in a high-level fusion architecture for cooperative perception systems,'' in
  \emph{2012 IEEE Intelligent Vehicles Symposium}, 2012.

\bibitem{kim2014multivehicle}
S.-W. Kim, B.~Qin, Z.~J. Chong, X.~Shen, W.~Liu, M.~H. Ang, E.~Frazzoli, and
  D.~Rus, ``Multivehicle cooperative driving using cooperative perception:
  Design and experimental validation,'' \emph{IEEE Transactions on Intelligent
  Transportation Systems}, vol.~16, no.~2, pp. 663--680, 2015.

\bibitem{etsi103324}
ETSI, ``{Intelligent Transport Systems (ITS); Cooperative Perception Services
  (CPS)},'' ETSI TS. 103 324, Early draft, 2020.

\bibitem{etsi-tr-103-562}
{ETSI}, ``{ETSI TR 103 562 - Intelligent Transport Systems (ITS); Vehicular
  Communications; Basic Set of Applications; Analysis of the Collective
  Perception Service (CPS)},'' Tech. Rep. V2.1.1, 2019.

\bibitem{qiu2018avr}
H.~Qiu, F.~Ahmad, F.~Bai, M.~Gruteser, and R.~Govindan, ``Avr: Augmented
  vehicular reality,'' in \emph{Proceedings of the 16th Annual International
  Conference on Mobile Systems, Applications, and Services}, 2018, pp. 81--95.

\bibitem{garlichs2019generation}
K.~Garlichs, H.-J. Günther, and L.~C. Wolf, ``Generation rules for the
  collective perception service,'' in \emph{2019 IEEE Vehicular Networking
  Conference (VNC)}, 2019, pp. 1--8.

\bibitem{zhou2018arve}
P.~Zhou, W.~Zhang, T.~Braud, P.~Hui, and J.~Kangasharju, ``Arve: Augmented
  reality applications in vehicle to edge networks,'' in \emph{Proceedings of
  the 2018 Workshop on Mobile Edge Communications}, 2018.

\bibitem{zhou2019enhanced}
------, ``Enhanced augmented reality applications in vehicle-to-edge
  networks,'' in \emph{22nd Conference on Innovation in Clouds, Internet and
  Networks and Workshops (ICIN)}.\hskip 1em plus 0.5em minus 0.4em\relax IEEE,
  2019.

\bibitem{a2w}
K.-Y. Lam, L.-H. Lee, and P.~Hui, ``A2w: Context-aware recommendation system
  for mobile augmented reality web browse,'' in \emph{The 29th ACM
  International Conference on Multimedia (ACM MM'21)}, 2021.

\bibitem{liu2020relevant}
W.~Liu, J.~Gori, O.~Rioul, M.~Beaudouin-Lafon, and Y.~Guiard, ``How relevant is
  hick's law for hci?'' in \emph{CHI}, 2020.

\bibitem{Schweickert1986ShorttermMC}
R.~Schweickert and B.~Boruff, ``Short-term memory capacity: magic number or
  magic spell?'' \emph{Journal of experimental psychology. Learning, memory,
  and cognition}, 1986.

\bibitem{thandavarayan2019analysis}
G.~Thandavarayan, M.~Sepulcre, and J.~Gozalvez, ``Analysis of message
  generation rules for collective perception in connected and automated
  driving,'' in \emph{2019 IEEE Intelligent Vehicles Symposium (IV)}, 2019.

\bibitem{ma2020artificial}
Y.~Ma, Z.~Wang, H.~Yang, and L.~Yang, ``Artificial intelligence applications in
  the development of autonomous vehicles: a survey,'' \emph{IEEE/CAA Journal of
  Automatica Sinica}, vol.~7, no.~2, pp. 315--329, 2020.

\bibitem{kebria2019deep}
P.~M. Kebria, A.~Khosravi, S.~M. Salaken, and S.~Nahavandi, ``Deep imitation
  learning for autonomous vehicles based on convolutional neural networks,''
  \emph{IEEE/CAA Journal of Automatica Sinica}, vol.~7, no.~1, pp. 82--95,
  2019.

\bibitem{cao2020rapid}
Z.~Cao, X.~Xu, B.~Hu, and M.~Zhou, ``Rapid detection of blind roads and
  crosswalks by using a lightweight semantic segmentation network,'' \emph{IEEE
  Transactions on Intelligent Transportation Systems}, 2020.

\bibitem{yoon2021performance}
D.~D. Yoon, B.~Ayalew, and G.~M.~N. Ali, ``Performance of decentralized
  cooperative perception in v2v connected traffic,'' \emph{IEEE Transactions on
  Intelligent Transportation Systems}, 2021.

\bibitem{kim2013cooperative}
S.-W. Kim, Z.~J. Chong, B.~Qin, X.~Shen, Z.~Cheng, W.~Liu, and M.~H. Ang,
  ``Cooperative perception for autonomous vehicle control on the road:
  Motivation and experimental results,'' in \emph{2013 IEEE/RSJ International
  Conference on Intelligent Robots and Systems}, 2013.

\bibitem{gunther2016collective}
H.-J. Günther, R.~Riebl, L.~Wolf, and C.~Facchi, ``Collective perception and
  decentralized congestion control in vehicular ad-hoc networks,'' in
  \emph{2016 IEEE Vehicular Networking Conference (VNC)}, 2016, pp. 1--8.

\bibitem{etsi-its-g5-5g}
ETSI, ``{Intelligent Transport Systems (ITS); ITS-G5 Access Layer Specification
  for Intelligent Transport Systems Operating in the 5 GHz Frequency Band},''
  ETSI TR. 302 663 V1.3.0, 2019.

\bibitem{kim2016cooperative}
S.-W. Kim and W.~Liu, ``Cooperative autonomous driving: A mirror neuron
  inspired intention awareness and cooperative perception approach,''
  \emph{IEEE Intelligent Transportation Systems Magazine}, vol.~8, no.~3, pp.
  23--32, 2016.

\bibitem{aoki2020}
S.~Aoki, T.~Higuchi, and O.~Altintas, ``Cooperative perception with deep
  reinforcement learning for connected vehicles,'' in \emph{2020 IEEE
  Intelligent Vehicles Symposium (IV)}.\hskip 1em plus 0.5em minus 0.4em\relax
  IEEE, pp. 328--334.

\bibitem{qiu2017augmented}
H.~Qiu, F.~Ahmad, R.~Govindan, M.~Gruteser, F.~Bai, and G.~Kar, ``Augmented
  vehicular reality: Enabling extended vision for future vehicles,'' in
  \emph{Proceedings of the 18th International Workshop on Mobile Computing
  Systems and Applications}, 2017, pp. 67--72.

\bibitem{chen2019cooper}
Q.~Chen, S.~Tang, Q.~Yang, and S.~Fu, ``Cooper: Cooperative perception for
  connected autonomous vehicles based on 3d point clouds,'' in \emph{2019 IEEE
  39th International Conference on Distributed Computing Systems (ICDCS)},
  2019, pp. 514--524.

\bibitem{wolcott2015fast}
R.~W. Wolcott and R.~M. Eustice, ``Fast lidar localization using
  multiresolution gaussian mixture maps,'' in \emph{2015 IEEE International
  Conference on Robotics and Automation (ICRA)}, 2015, pp. 2814--2821.

\bibitem{hata2015feature}
A.~Y. Hata and D.~F. Wolf, ``Feature detection for vehicle localization in
  urban environments using a multilayer lidar,'' \emph{IEEE Transactions on
  Intelligent Transportation Systems}, vol.~17, no.~2, pp. 420--429, 2016.

\bibitem{glenn_jocher_2020_3983579}
\BIBentryALTinterwordspacing
G.~J. et~al., ``ultralytics/yolov5: v3.0,'' 2020. [Online]. Available:
  \url{https://doi.org/10.5281/zenodo.3983579}
\BIBentrySTDinterwordspacing

\bibitem{saej2735}
{SAE International}, ``V2x communications message set dictionary,'' 2020.

\bibitem{friedner20165g}
L.~telcom, ``In-car mobile signal attenuation measurements,'' \emph{LS telcom
  UK, Tech. Rep}, 2017.

\bibitem{noor2018performance}
M.~Noor-A-Rahim, G.~M.~N. Ali, H.~Nguyen, and Y.~L. Guan, ``Performance
  analysis of ieee 802.11 p safety message broadcast with and without relaying
  at road intersection,'' \emph{IEEE Access}, vol.~6, pp. 23\,786--23\,799,
  2018.

\bibitem{tonguz2010dv}
O.~K. Tonguz, N.~Wisitpongphan, and F.~Bai, ``Dv-cast: A distributed vehicular
  broadcast protocol for vehicular ad hoc networks,'' \emph{IEEE Wireless
  Communications}, vol.~17, no.~2, pp. 47--57, 2010.

\bibitem{karp2000gpsr}
B.~Karp and H.-T. Kung, ``Gpsr: Greedy perimeter stateless routing for wireless
  networks,'' in \emph{Proceedings of the 6th annual international conference
  on Mobile computing and networking}, 2000, pp. 243--254.

\bibitem{lochert2005geographic}
C.~Lochert, M.~Mauve, H.~F{\"u}{\ss}ler, and H.~Hartenstein, ``Geographic
  routing in city scenarios,'' \emph{ACM SIGMOBILE mobile computing and
  communications review}, vol.~9, no.~1, pp. 69--72, 2005.

\bibitem{granelli2007enhanced}
F.~Granelli, G.~Boato, D.~Kliazovich, and G.~Vernazza, ``Enhanced gpsr routing
  in multi-hop vehicular communications through movement awareness,''
  \emph{IEEE Communications Letters}, vol.~11, no.~10, pp. 781--783, 2007.

\bibitem{gsma-cv2x}
GSMA, ``{Cellular Vehicle-to-everything; Enabling intelligent transport}.''

\bibitem{peden2004}
M.~Peden, R.~Scurfield, D.~Sleet, A.~A. Hyder, C.~Mathers, E.~Jarawan,
  A.~Hyder, D.~Mohan, and E.~Jarawan, \emph{World report on road traffic injury
  prevention}.\hskip 1em plus 0.5em minus 0.4em\relax World Health
  Organization, 2004.

\bibitem{dirnbach2020methodology}
I.~Dirnbach, T.~Kubjatko, E.~Kolla, J.~Ondru{\v{s}}, and
  {\v{Z}}.~{\v{S}}ari{\'c}, ``Methodology designed to evaluate accidents at
  intersection crossings with respect to forensic purposes and transport
  sustainability,'' \emph{Sustainability}, 2020.

\bibitem{higuchi2019value}
T.~Higuchi, M.~Giordani, A.~Zanella, M.~Zorzi, and O.~Altintas,
  ``Value-anticipating v2v communications for cooperative perception,'' in
  \emph{2019 IEEE Intelligent Vehicles Symposium (IV)}.\hskip 1em plus 0.5em
  minus 0.4em\relax IEEE, 2019.

\bibitem{favaro2017examining}
F.~M. Favar{\`o}, N.~Nader, S.~O. Eurich, M.~Tripp, and N.~Varadaraju,
  ``Examining accident reports involving autonomous vehicles in california,''
  \emph{PLoS one}, 2017.

\bibitem{caesar2020nuscenes}
H.~Caesar, V.~Bankiti, A.~H. Lang, S.~Vora, V.~E. Liong, Q.~Xu, A.~Krishnan,
  Y.~Pan, G.~Baldan, and O.~Beijbom, ``nuscenes: A multimodal dataset for
  autonomous driving,'' in \emph{CVPR}, 2020.

\bibitem{hannah2020long}
B.~Hannah~Topliss, C.~Harvey, and G.~Burnett, ``How long can a driver look?
  exploring time thresholds to evaluate head-up display imagery,'' in
  \emph{AutomotiveUI}, 2020.

\bibitem{thandavarayan2020generation}
G.~Thandavarayan, M.~Sepulcre, and J.~Gozalvez, ``Generation of cooperative
  perception messages for connected and automated vehicles,'' \emph{IEEE
  Transactions on Vehicular Technology}, vol.~69, no.~12, pp. 16\,336--16\,341,
  2020.

\bibitem{thandavarayan2020redundancy}
G.~Thandavarayan\vspace{0mm}, M.~Sepulcre, and J.~Gozalvez, ``Redundancy
  mitigation in cooperative perception for connected and automated vehicles,''
  in \emph{2020 IEEE 91st Vehicular Technology Conference}, 2020.

\bibitem{corneo2019age}
L.~Corneo, C.~Rohner, and P.~Gunningberg, ``Age of information-aware scheduling
  for timely and scalable internet of things applications,'' in \emph{IEEE
  Conference on Computer Communications}, 2019, pp. 2476--2484.

\bibitem{correia2017avui}
N.~N. Correia and A.~Tanaka, ``Avui: Designing a toolkit for audiovisual
  interfaces,'' in \emph{CHI}, 2017.

\bibitem{papadakis2020blocking}
G.~Papadakis, D.~Skoutas, E.~Thanos, and T.~Palpanas, ``Blocking and filtering
  techniques for entity resolution: A survey,'' \emph{ACM Computing Surveys
  (CSUR)}, 2020.

\bibitem{cheng2018big}
N.~Cheng, F.~Lyu, J.~Chen, W.~Xu, H.~Zhou, S.~Zhang, and X.~Shen, ``Big data
  driven vehicular networks,'' \emph{IEEE Network}, vol.~32, no.~6, pp.
  160--167, 2018.

\bibitem{sun2018analyzing}
D.~Sun, K.~Zhang, and S.~Shen, ``Analyzing spatiotemporal traffic line source
  emissions based on massive didi online car-hailing service data,''
  \emph{Transportation Research Part D: Transport and Environment}, vol.~62,
  pp. 699--714, 2018.

\bibitem{silva2016survey}
N.~F. F.~D. Silva, L.~F. Coletta, and E.~R. Hruschka, ``A survey and
  comparative study of tweet sentiment analysis via semi-supervised learning,''
  \emph{ACM Computing Surveys (CSUR)}, vol.~49, no.~1, pp. 1--26, 2016.

\bibitem{triguero2015self}
I.~Triguero, S.~Garc{\'\i}a, and F.~Herrera, ``Self-labeled techniques for
  semi-supervised learning: taxonomy, software and empirical study,''
  \emph{Knowledge and Information systems}, vol.~42, no.~2, pp. 245--284, 2015.

\bibitem{de2000mahalanobis}
R.~Maesschalck, D.~Jouan-Rimbaud, and D.~Massart, ``The mahalanobis distance,''
  \emph{Chemometrics and Intelligent Laboratory Systems}, vol.~50, pp. 1--18,
  2000.

\bibitem{weinberger2009distance}
K.~Q. Weinberger and L.~K. Saul, ``Distance metric learning for large margin
  nearest neighbor classification,'' \emph{J. Mach. Learn. Res.}, vol.~10, p.
  207–244, Jun. 2009.

\bibitem{kulis2013metric}
B.~Kulis \emph{et~al.}, ``Metric learning: A survey,'' \emph{Foundations and
  Trends in Machine Learning}, 2013.

\bibitem{9163287}
P.~Zhou, T.~Braud, A.~Zavodovski, Z.~Liu, X.~Chen, P.~Hui, and J.~Kangasharju,
  ``Edge-facilitated augmented vision in vehicle-to-everything networks,''
  \emph{IEEE Transactions on Vehicular Technology}, vol.~69, no.~10, pp.
  12\,187--12\,201, 2020.

\bibitem{bochinski2017high}
E.~Bochinski, V.~Eiselein, and T.~Sikora, ``High-speed tracking-by-detection
  without using image information,'' in \emph{AVSS}, 2017.

\bibitem{de2020metric}
W.~De~Vazelhes, C.~Carey, Y.~Tang, N.~Vauquier, and A.~Bellet, ``metric-learn:
  Metric learning algorithms in python,'' \emph{Journal of Machine Learning
  Research}, 2020.

\bibitem{sommer2011bidirectionally}
C.~Sommer, R.~German, and F.~Dressler, ``{Bidirectionally Coupled Network and
  Road Traffic Simulation for Improved IVC Analysis},'' \emph{TMC}, 2011.

\bibitem{sommer2011using}
C.~Sommer and F.~Dressler, ``Using the right two-ray model ? a
  measurement-based evaluation of phy models in vanets,'' 2011.

\bibitem{sommer2013ivc}
C.~Sommer, D.~Eckhoff, and F.~Dressler, ``Ivc in cities: Signal attenuation by
  buildings and how parked cars can improve the situation,'' \emph{IEEE
  Transactions on Mobile Computing}, vol.~13, no.~8, pp. 1733--1745, 2014.

\bibitem{code}
\BIBentryALTinterwordspacing
P.~Zhou, ``{AICP} simulation code and datasets,'' 2021. [Online]. Available:
  \url{https://github.com/pengyuan-zhou/AICP}
\BIBentrySTDinterwordspacing

\bibitem{kornek2010effects}
D.~Kornek, M.~Schack, E.~Slottke, O.~Klemp, I.~Rolfes, and T.~Kürner,
  ``Effects of antenna characteristics and placements on a vehicle-to-vehicle
  channel scenario,'' in \emph{2010 IEEE International Conference on
  Communications Workshops}, 2010, pp. 1--5.

\bibitem{eckhoff2016impact}
D.~Eckhoff, A.~Brummer, and C.~Sommer, ``On the impact of antenna patterns on
  vanet simulation,'' in \emph{2016 IEEE Vehicular Networking Conference
  (VNC)}, 2016, pp. 1--4.

\bibitem{colley2017design}
A.~Colley, J.~H{\"a}kkil{\"a}, B.~Pfleging, and F.~Alt, ``A design space for
  external displays on cars,'' in \emph{AutomotiveUI}, 2017.

\bibitem{lateral-position-5444759}
S.~{Onimaru} and M.~{Kitazaki}, ``Visual and tactile information to improve
  drivers' performance,'' in \emph{IEEE Virtual Reality Conference}, 2010.

\bibitem{onimaru2008cross}
S.~Onimaru, T.~Uraoka, N.~Matsuzaki, and M.~Kitazaki, ``Cross-modal information
  display to improve driving performance,'' in \emph{VRST}, 2008.

\bibitem{hauslschmid2015augmenting}
R.~H{\"a}uslschmid, S.~Osterwald, M.~Lang, and A.~Butz, ``Augmenting the
  driver's view with peripheral information on a windshield display,'' in
  \emph{IUI}, 2015.

\bibitem{chaturvedi2019peripheral}
I.~Chaturvedi, F.~H. Bijarbooneh, T.~Braud, and P.~Hui, ``Peripheral vision: a
  new killer app for smart glasses,'' in \emph{IUI}, 2019.

\end{thebibliography}
\end{document}